\documentclass[11pt]{article}
\usepackage{amsmath}
\usepackage{amssymb}
\usepackage{latexsym}

\newcommand{\st}[1]{\ensuremath{^{\scriptstyle \textrm{#1}}}}

\newcommand{\vac}{|0\rangle}

\newcommand\bigcheck[1]{#1 \raise1ex\hbox{$\hspace{-1ex}{}^\vee$}}
\newcommand\sucheck[1]{#1 \raise0.5ex\hbox{$\hspace{-1ex}{}^\vee$}}

\newcommand{\alphaparenlist}{
  \renewcommand{\theenumi}{\alph{enumi}}%
  \renewcommand{\labelenumi}{(\theenumi)}%
}

\newcommand{\Alist}{
  \renewcommand{\theenumi}{\arabic{enumi}}%
  \renewcommand{\labelenumi}{(A\theenumi)}%
}

\newcommand{\ad}{\mathop{\rm ad}}
\newcommand{\C}{\mathcal{C}}
\newcommand{\CC}{\hbox{\bf C}}
\newcommand{\QQ}{\hbox{\bf Q}}
\newcommand{\RR}{\hbox{\bf R}}

\renewcommand{\hat}{\widehat}

\normalbaselines

\begin{document}
\bibliographystyle{unsrt}

\newtheorem{theorem}{Theorem}[section]
\newtheorem{lemma}{Lemma}[section]
\newtheorem{proposition}{Proposition}[section]
\newtheorem{corollary}{Corollary}[section]

\renewcommand{\theequation}{\thesection.\arabic{equation}}

\def\bea*{\begin{eqnarray*}}
\def\eea*{\end{eqnarray*}}
\def\ba{\begin{array}}
\def\ea{\end{array}}
\count1=1
\def\be{\ifnum \count1=0 $$ \else \begin{equation}
\fi}
\def\ee{\ifnum\count1=0 $$ \else
\end{equation}\fi}
\def\ele(#1){\ifnum\count1=0 \eqno({\bf #1}) $$
\else \label{#1}\end{equation}\fi}
\def\req(#1){\ifnum\count1=0 {\bf #1}\else
\ref{#1}\fi}
\def\bea(#1){\ifnum \count1=0   $$
\begin{array}{#1}
\else \begin{equation} \begin{array}{#1} \fi}
\def\eea{\ifnum \count1=0 \end{array} $$
\else  \end{array}\end{equation}\fi}
\def\elea(#1){\ifnum \count1=0 \end{array}
\label{#1}\eqno({\bf #1}) $$
\else\end{array}\label{#1}\end{equation}\fi}
\def\cit(#1){
\ifnum\count1=0 {\bf #1} \cite{#1} \else
\cite{#1}\fi}
\def\bibit(#1){\ifnum\count1=0 \bibitem{#1} [#1    ] \else \bibitem{#1}\fi}
\def\ds{\displaystyle}
\def\hb{\hfill\break}
\def\comment#1{\hb {***** {\em #1} *****}\hb }

\newcommand{\F}{\mathcal{F}}

\newcommand{\TZ}{\hbox{\bf T}}
\newcommand{\MZ}{\hbox{\bf M}}
\newcommand{\ZZ}{\hbox{\bf Z}}
\newcommand{\NZ}{\hbox{\bf N}}
\newcommand{\RZ}{\hbox{\bf R}}
\newcommand{\CZ}{\,\hbox{\bf C}}
\newcommand{\PZ}{\hbox{\bf P}}
\newcommand{\QZ}{\hbox{\bf Q}}
\newcommand{\HZ}{\hbox{\bf H}}
\newcommand{\EZ}{\hbox{\bf E}}
\newcommand{\GZ}{\,\hbox{\bf G}}

\font\germ=eufm10
\def\goth#1{\hbox{\germ #1}}
\vbox{\vspace{38mm}}

\begin{center}
{\LARGE \bf Quantum Reduction for
Affine Superalgebras}\\[5mm]

Victor Kac \\{\it Department of
Mathematics,  M. I. T. \\
Cambridge, MA 02139, USA \\
 (e-mail: kac@math.mit.edu )}
\\ [2mm]
Shi-shyr Roan \\ {\it
Institute of Mathematics
, Academia Sinica \\  Taipei , Taiwan \\ (e-mail:
maroan@ccvax.sinica.edu.tw)} \\ [2mm]
Minoru Wakimoto \\{\it Graduate School of
Mathematics , Kyushu University \\
Fukuoka, 812-8581, Japan \\
 (e-mail:wakimoto@math.kyushu-u.ac.jp)}
\\ [5mm]
\end{center}

\begin{abstract}
We extend the homological method of  quantization of
generalized Drinfeld--Sokolov reductions to affine
superalgebras.  This leads, in particular, to a unified
representation theory of superconformal algebras.

\par \vspace{5mm} \noindent
 \par \noindent
1990 PACS: 02.20, 03.65.F, 89

\end{abstract}

\setcounter{section}{-1}
\section{Introduction}
\setcounter{equation}{0}

A series of papers on $W$-algebras written in the second half of the 1980's and the early 1990's
(see \cite{BS}) culminated in the work of Feigin and Frenkel \cite{FF1, FF2} who showed that to a simple finite-dimensional Lie algebra ${\goth g}$ one canonically associates a $W$-algebra $W_k(\goth g)$ as a result of quantization of the classical Drinfeld--Sokolov reduction.  Namely, $W_k (\goth g)$ is realized as homology of a BRST complex involving the principal nilpotent element of $\goth g$ (i.e.,~the nilpotent element the closure of whose orbit contains all other nilpotent elements), the universal enveloping algebra of the affine Kac--Moody algebra $\hat{\goth g}$ associated to $\goth g$, and the charged fermionic ghosts associated to the currents of a maximal nilpotent subalgebra $\goth n$ of $\goth g$.

This approach allows one not only to define the $W$-algebras, but also to construct a functor $H$ from the category of restricted $\hat{\goth g}$-modules of level $k$ to the category of positive energy modules over $W_k (\goth g)$.  Namely, the $W_k(\goth g)$-module corresponding to a $\hat{\goth g}$-module is the homology $H(M)$ of the BRST complex associated to $M$.  This functor was applied in \cite{FKW} to the admissible $\hat{\goth g}$-modules, classified in \cite{KW1}, \cite{KW2}, in order to compute the characters of $W_k (\goth g)$-modules. (In the simplest case of $\goth g =s\ell_2$ one recovers thereby the minimal series modules over the Virasoro algebra $= W_k (s \ell_2)$.)

It is straightforward to generalize this construction to the case when
$f$ is an even nilpotent element, that is for the $s\ell_2$-triple
$\langle e,x,f \rangle$, such that $[e,f]=x$, $[x,e]=e$, $[x,f]=-f$, all
eigenvalues of ${\rm ad}\,\, x$ are integers (for general $f$ they lie
in $\frac{1}{2} \ZZ$).  One just takes instead of $\goth n$ the
subalgebra $\goth g_+$ of $\goth g$ spanned by eigenspaces with
positive eigenvalues for ${\rm ad} \,\, x$.  Unfortunately, most
nilpotent elements are not even, but often one can replace $x$ by $x'$
such that ${\rm ad} \,\,x'$ has integer eigenvalues, so that the
construction gives the same homology (see e.g.~\cite{BT}).  However,
it remained unclear how to make it work for a general simple Lie
algebra $\goth g$ and a general nilpotent element $f$.  The situation
gets worse if one tries to go to the Lie superalgebra case since
already the simplest Lie superalgebra $spo (2|1)$ has no good $\ZZ$-
gradations.

In the present paper we show how to resolve this problem.  It turns out that one needs only to add neutral fermionic ghosts associated to the currents of the eigenspace $\goth g_{1/2}$ of ${\rm ad}\,\,x$.

This is done in Section~2, where to each quadruple $(\goth g ,x,f,k)$, where $\goth g$ is a simple finite-dimensional Lie superalgebra with a fixed even invariant bilinear form $(.|.)$, $x$ is an ${\rm ad}$-diagonizable element of $\goth g$ with eigenvalues in $\frac{1}{2}\ZZ$, $f$ is a nilpotent even element of $\goth g$ such that $[x,f]=-f$, and $k \in \CC$, we associate a BRST complex
\begin{displaymath}
  ({\cal C} (\goth g ,x,f,k) =V_k (\goth g) \otimes F^{\rm ch}
     \otimes F^{\rm ne} \, , d_0) \, .
\end{displaymath}
Here $V_k (\goth g)$ is the universal affine vertex algebra of level $k$
associated to $\hat{\goth g}$, $F^{\rm ch}$ is the vertex algebra of
free charged fermions based on $\goth g_+ + \goth g^*_+$ with
reversed parity, $F^{\rm ne}$ is the vertex algebra of free neutral
fermions based on $\goth g_{1/2}$ with the form $\langle a,b \rangle =
(f|[a,b])$, and $d_0$ is an explicitly constructed odd derivation of
the vertex algebra ${\cal C} (\goth g ,x,f,k)$ whose square is $0$ (see
Section~2.1).  The main object of our study is the $0$\st{th}
homology of this complex, which is a vertex algebra, denoted by $W_k
(\goth g ,x,f)$.  In the case when the pair $(x,f)$ can be included in
an $s\ell_2$-triple $(e,x,f)$ (then $x$ is determined by $f$ up to
conjugation), we denote this vertex algebra by $W_k (\goth g ,f)$.  In
this case the map ${\rm ad} f: \goth g_{1/2} \to \goth g_{-1/2}$ is
an isomorphism, which suffices for the construction of the
energy-momentum field $L(z)$ of $W_k (\goth g ,x,f)$ (see
Section~2.2); under the same assumption, we construct fields $J^{\{ v
  \}}$ in $W_k (\goth g ,x,f)$ of conformal weight $1$, corresponding
to each element $v \in \goth g^{x,f} $, the centralizer of $x$ and $f$
(see Section~2.4).

As in \cite{FF2, FKW}, given a restricted $\hat{\goth g}$-module $M$
of level $k$, hence a $V_k (\goth g)$-module, we extend it to a ${\cal C}
(\goth g ,x,f,k)$-module ${\cal C} (M) = M \otimes
F^{\rm ch} \otimes F^{\rm ne}$, which gives rise to a complex (${\cal C}
(M) \, , \, d_0$) of ${\cal C} (\goth g ,x,f,k)$-modules.  Its homology
$H(M)$ is a $W_k (\goth g ,x,f)$-module.  In Section~3.1 we compute
the Euler--Poincar\'e character of this module:
\begin{displaymath}
  {\rm ch}_{H(M)} (h) = \sum_{j \in \ZZ} (-1)^j
  {\rm tr}_{H_j(M)} q^{L_0} e^{2\pi i J^{\{ h \}}_0} \, ,
\end{displaymath}
where $h$ is an element of a Cartan subalgebra of $\goth g^{x,f}$ and $J^{\{ h \}}$ is the corresponding field of $W_k ({\goth g},x,f)$.
Furthermore, in Section~3.2 we find necessary and sufficient
conditions on the $\hat{\goth g}$-module $M$ for the non-vanishing of
${\rm ch}_{H(M)}$.  The $\hat{\goth g}$-modules $M$ satisfying these
conditions are called non-degenerate.

In Section~3.3 we recall the definition of admissible highest weight
$\hat{\goth g}$-modules $L(\Lambda)$ in the Lie superalgebra case
\cite{KW4}.  The characters of these modules in the Lie algebra case
were computed in \cite{KW1}.  Unfortunately we do not know how to
prove an analogous character formula even in its weaker form in the Lie superalgebra case.  This character formula is our first fundamental
conjecture (which is confirmed by many examples in \cite{KW1}, \cite{KW2},
  \cite{KW4}).  The second fundamental conjecture states that the $W_k
(\goth g ,x,f)$-module $H(M)$ is either zero or irreducible, provided
that $(x,f)$ is a ``good'' pair and $M$ is an admissible highest
weight $\hat{\goth g}$-module.  Of course, these conjectures allow us to
compute the characters of irreducible $W_k (\goth g ,x,f)$-modules
$H(M)$ for non-degenerate admissible $\hat{\goth g}$-modules, using the
results of Section~3.1.

In Section~4 we study the vertex algebra $W_k (\goth g ,f)$ in the case a ``minimal'' nilpotent even element $f$, namely when $f$ is a root vector corresponding to an even highest root of $\goth g$.  These vertex algebras were considered from a quite different viewpoint in \cite{FL}, and they include all well known superconformal algebras, like the $N\leq 4$ superconformal algebras and the big $N=4$ superconformal algebras.

In Section~5 we show (following  \cite{FKW}) that indeed
all non-degenerate admissible $\hat{s \ell}_2$-modules produce all
minimal series Virasoro modules via the functor $M \to H(M)$.
In Section~6 we show, in a similar fashion, that all non-degenerate
admissible $spo (2|1)^{\hat{}}$-modules (whose characters were
computed in \cite{KW1} as well) produce all characters of minimal series
Neveu--Schwarz modules.  Finally, in Section~7, using the conjectural
character formulas for ``boundary'' admissible $s \ell
(2|1)^{\hat{}}$-modules, we recover the characters of all minimal
series modules over the $N=2$ superconformal algebra.  Note that it
was already established by Khovanova \cite{Kh} that the classical
reduction of $s\ell (2|1)^{\hat{}}$ produces the $N=2$ superconformal
algebra.

Further examples and results  are presented in \cite{KW5}, where, in particular,
we give a proof of a stronger form of the fundamental Conjecture 2.1
of the present paper.

The results of this paper were reported at the ICM in Beijing \cite{K5}.

Throughout the paper all vector spaces, algebras and tensor products are considered over the field of complex numbers $\CC$, unless otherwise stated.  We denote by $\ZZ$, $\QQ$ and $\RR$ the rings of integers, rational and real numbers, respectively, and by $\ZZ_+$ the set of non-negative integers..

\section{An Overview of the Operator Product Expansion}
\setcounter{equation}{0}
In this section, we give
a brief summary of  some basic properties of the
operator product expansion (OPE) which will be used in this paper (for the details, see
\cite{K4} or \cite{W}).

Let $A$ be a Lie superalgebra with a
central element $K$ and a $\ZZ$-filtration by
subspaces,
$$
\cdots \supset  A_{(0)} \supset A_{(1)} \supset A_{(2)}
\supset \cdots
$$
where $ \bigcup_j A_{(j)}=A \, , \, \bigcap_j A_{(j)} = 0$ and
$[ A_{(i)} , A_{(j)} ] \subset A_{(i+j)}$.  Throughout this
paper, we always write $[ \ , \ ]$ for the
Lie superbracket. For a given complex number $k
\in \CZ$, we denote by $U_k (A)$ the
quotient of the universal enveloping algebra of
$A$ by the ideal generated by $K - k
\cdot 1$, and by $U_k (A)^{\rm com}$ the
completion of $U_k (A)$, which consists
of all series $\sum_j u_j \ (u_j \in U_k (A) )$,
such that for each
$N \in \ZZ_+$ all but a finite number of
the $u_j$'s lie in $U_k (A) A_{(N)}$.
Then $U_k (A)^{\rm com}$ is an
associative algebra containing $U_k (A)$.
Any $A$-module $M$ in which every element
of $M$ is annihilated by some $A_{(N)}$, can
be uniquely extended to a module over $U_k
(A)^{\rm com}$. Such a module over
$A$ is called a {\it restricted} $A$-module.

A $U_k (A)^{\rm com}$-valued {\it field} is
an expression of the form
$$
a (z) =  \sum_{n \in  \ZZ} a_{(n)} z^{-n-1} \ ,
$$
where $a_{(n)} \in U_k (A)^{\rm com}$
satisfy the property that for each $N \in \ZZ_+$,
$a_{(n)} \in U_k (A)^{\rm com}A_{(N)}$ for $n \gg 0$,
and all $a_{(n)}$ have the same parity, which
will be denoted by $p ( a )
\in \ZZ/2 \ZZ$. Note that for a restricted $A$-module $M$,
the image of a field in ${\rm End} (M)$ gives
rise to a usual ${\rm End} (M)$-valued field.
It is easy to see that the derivative $\partial_z
a (z)$ of a field $a ( z )$ is also a field. The
{\it normal ordered product} of two fields $a (z)$
and
$ b (z)$ is defined by
$$
: a(z) b(z): = a(z)_- b(z) + (-1)^{p(a) p(b)} b(z) a(z)_+ \ ,
$$
where $a(z)_+ = \sum_{n < 0} a_{(n)} z^{-n-1}$ and $a(z)_- = \sum_{n \geq 0} a_{(n)} z^{-n-1}$.
For $n \in \ZZ$, the {\it $n$-th product}
$a(z)_{(n)}b(z) ( = (a_{(n)}b)(z)) $ of $a (z)$
and $ b (z)$ is defined as follows. For a
non-negative integer $n$,
$$
a(z)_{(n)}b(z) = {\rm Res}_x ( x - z)^n
[ a (x) , b (z) ] \ ,
$$
and
$$
a(z)_{(-n-1)}b(z) = \frac{: \partial_z^n a(z)b(z):}{n ! } \ .
$$
The $n$\st{th} products of fields $a (z)$ and $ b
(z)$  for $n \in \ZZ_+$ are encoded in the
{\it $\lambda$-bracket} defined by
$$
[ a_\lambda b] = \sum_{n \in \ZZ_+} \frac{\lambda^n}{n !} a_{(n)}b \ ,
$$
which is in general a formal power series in
$\lambda$ ( with coefficients in $U_k (A)^{\rm
com}$ ). Here and further on, we often drop
the indeterminate $z$, e.g., we shall write
$\partial a$ in place of $\partial_z a (z)$.
\begin{proposition}  \label{prop:lambda}
$\cite{K4}$   The following properties hold for
the $\lambda$-bracket:
$$
\begin{array}{cl}
( {\rm sesquilinearity} ) & [ \partial a_\lambda b]
= - \lambda [a_\lambda b] , \ \, \ \, \ \ \ \ [  a_\lambda \partial b]
= (\partial + \lambda) [a_\lambda b] \ ; \\
({\rm Jacobi \ identity}) & [a_\lambda
[b_\mu c] ]  = [ [a_\lambda b]_{\lambda+\mu} c] +
(-1)^{p(a) p(b)}[ b_\mu [a_\lambda c]] \ ;
\\
({\rm noncommutative \ Wick \ formula}) &
[a_\lambda :bc:] =:[a_\lambda b]c:
+(-1)^{p(a) p(b)} :b[a_\lambda c]:+
\int_{0}^{\lambda}  [[a_\lambda b]_\mu c] d \mu \ .
\end{array}
$$
\end{proposition}
Recall that a pair $(a(z), b(z))$ of fields  is
called {\it local} if
$$
(z - w )^N [ a (z), b (w) ] = 0 \ , \ \ {\rm for} \ N \gg 0 \ .
$$
Note that the $\lambda$-bracket
 of two local fields is a polynomial in $\lambda$.
\begin{proposition}  \label{prop:lamloc}
$\cite{K4}$  Let $(a(z), b(z))$ be a local pair
of fields. Then

{\rm (a)}
\be
[a_{(m)}, b_{(n)}] = \sum_{j \in
\ZZ_+}{  m
 \choose j} (a_{(j)}b)_{(m+n-j)} .
\ele(cmr)

{\rm (b)} The $\lambda$-bracket satisfies the
properties:
$$
\begin{array}{cl}
({\rm skewcommutativity}) &  [ a_\lambda b] =
- (-1)^{p(a) p(b)} [b_{-\lambda- \partial} a]  \
;
\\
({\rm right \  noncommutative \ Wick \ formula })
\cite{BK} & [:ab:_\lambda c]  = :(e^{\partial
\frac{d}{d
\lambda}} a)[b_\lambda c]: +(-1)^{p(a)
p(b)}:(e^{\partial
\frac{d}{d
\lambda}} b)[a_\lambda c]: \\
& \ \ \ \ \ \ \ \ +(-1)^{p(a) p(b)}
\int_{0}^{\lambda}  [b_\mu [a_{\lambda- \mu} c]] d
\mu \ .
\end{array}
$$

{\rm (c)} The normal order commutator of $a(z)$
and
$b(z)$ is expressed via the $\lambda$-bracket:
\be
:ab: - (-1)^{p(a) p(b)} :ba: \ \ =  \int_{-\partial}^0
[a_\lambda b] d \lambda \ .
\ele(opcom)
\end{proposition}

Note that formula (\req(cmr)) is nothing else but
(the singular part of) the operator product
expansion (OPE) for the local pair $(a(z), b(z))$:
$$
[a(z) , b(w) ] \ = \ \sum_{j = 0}^N
\frac{\partial_w^j \delta (z-w)}{j !}
a(w)_{(j)}b(w)
$$
where $\delta(z-w) = z^{-1} \sum_{n \in \ZZ}
(\frac{w}{z})^n$ is the formal $\delta$-function.
Propositions \ref{prop:lambda} and
\ref{prop:lamloc} provide an efficient and
convenient way of calculating the OPE of local
pairs.

Of course, in the case of normal ordered products of any number
of free fields, one can use the usual Wick formula (see
e.g.~\cite{K4}).
Note that  (\req(cmr)) immediately
implies the following corollary.
\begin{corollary} \label{cor:a0b0}
If $(a(z), b(z))$ is a local pair with $a(z)_{(0)}b(z)= 0$, then $[a_{(0)}, b(z)] = 0$.
\end{corollary}

Given a collection ${\cal V}$ of pairwise local
($U_k (A)^{\rm com}$-valued) fields, we
may consider its closure $V = V_k ( A,
{\cal V})$ which is the minimal space of fields
containing $1$ and ${\cal V}$, closed under
$\partial_z$ and all
$n$-th products ( $n \in \ZZ$ ). By Dong's lemma
\cite{K4}, $V$ consists of pairwise local
fields, hence Propositions \ref{prop:lambda} and
\ref{prop:lamloc} also apply to fields in $V$.
Note that $V$ is a vertex algebra and any
restricted $A$-module $M$ extends uniquely to a
$V$-module.

{\bf Example 1.1} (energy-momentum field). Let
$Vir$ be the Virasoro algebra, i.e.,~the Lie
algebra with the basis $L_j \ (j \in \ZZ)$ and a
central element $C$, with the commutation
relations
$$
[ L_m , L_n ] = (m-n) L_{m+n} + \delta_{m, -n}
\frac{(m^3-m)C}{12} \ .
$$
We take the filtration $Vir_{(j)}=\CC C +\sum_{i \geq j} \CC L_j$
for $j\leq 0$, $Vir_{(j)} = \sum_{i \geq j}
\CZ L_j$ for $j>0$.  Let $L (z) = \sum_{n
\in \ZZ} L_n z^{-n-2}$ (note that $L_{n} =
L_{(n+1)}$). This field is local with itself, so
that the commutation relations of $L_j$'s are
encoded by the $\lambda$-bracket,
\be
[ L_\lambda L ] = ( \partial + 2 \lambda ) L + \frac{\lambda^3 c}{12} \ .
\ele(Vir)
Here $c \in \CZ$ is
the eigenvalue of $C$.

A local field $L(z)$ with
the $\lambda$-bracket (\req(Vir)) is called an
{\it energy-momentum field} with {\it central
charge} $c$.

Fix an energy-momentum field
$L = L(z)$. Let $a (z)$ be a field such that $(L,
a)$ is a local pair. One says that the field $a$
has {\it conformal weight} $\triangle \in \CZ$
(with respect to $L$) if the following relation
holds:
$$
[  L_\lambda a ] = ( \partial + \triangle \lambda ) a + o (\lambda) \ .
$$
Note that in this case $\partial_z a(z)
(= \partial a)$ has conformal weight
$\triangle + 1$. In the special case, when $[
L_\lambda a ] = ( \partial + \triangle \lambda )
a$, one calls $a$ a {\it primary} field. When
$a(z)$ is a field with conformal weight
$\triangle$, it is convenient to change the
indexation of the modes of $a (z)$ :
$$
a (z)  = \sum_{n \in  \ZZ}
a_{(n)} z^{-n-1} = \sum_{n \in -\triangle + \ZZ}
a_n z^{-n-\triangle} \ , \ \ a_n = a_{(n +
\triangle -1)} \ .
$$
For example $L(z)$ has the conformal
weight $2$, and we write $L(z) = \sum_{n \in \ZZ}
L_n z^{-n-2}$.
\begin{proposition}  \label{prop:lOPE}
Let $a (z), b (z)$ be fields of conformal
weights $\triangle_a$ and $\triangle_b$
respectively. Then

{\rm (a) } $\triangle_{a_{(n)}b} = \triangle_a +
\triangle_b - n - 1$; in particular,
$\triangle_{:ab:} = \triangle_a + \triangle_b $.

{\rm (b)} The commutator formula $(\req(cmr))$
takes the homogeneous form:
$$
[a_m, b_n] = \sum_{j \in \ZZ_+} { \triangle_a + m
-1 \choose j}(a_{(j)}b)_{m+n} \ .
$$
\end{proposition}

Recall that a vector superspace is a vector space $V$ decomposed into a
direct sum of vector spaces $V_{\bar{0}}$ and $V_{\bar{1}}$
$(\bar{0}, \bar{1} \in \ZZ /2 \ZZ)$, called the \emph{even} and \emph{odd} part
of $V$, respectively.  We write $p(v)=\alpha$ if $v \in
V_{\alpha}$.  Denoting by $\Gamma$ the endomorphism of $V$ that
acts as $(-1)^{\alpha}$ on $V_{\alpha}$, we may define the
supertrace of $a \in {\rm End} V$ (provided that ${\rm dim} V <\infty$) by
\cite{K1}
\begin{displaymath}
 {\rm str}_V a = {\rm tr}_V (\Gamma a) \, .
\end{displaymath}
In particular, letting ${\rm sdim} V = {\rm str}_V I_V$, we have
${\rm sdim} V = {\rm dim} V_{\bar{0}} -{\rm dim} V_{\bar{1}}$.

Recall that a vertex algebra is called \emph{strongly generated}
by a collection of fields $\F$ if normally ordered products of
fields from $\CC [\partial] \F$ span the space of fields of this
vertex algebra.

{\bf Example 1.2} (neutral free superfermions).
Let $A = A_{\bar{0}} \oplus
A_{\bar{1}}$ be a finite-dimensional
superspace with a non-degenerate
skew-supersymmetric bilinear form $\langle . , .
\rangle$, i.e.,~$\langle A_{\bar{0}},
A_{\bar{1}} \rangle = 0 $ and $\langle . ,
.
\rangle$ is skewsymmetric (resp. symmetric ) on
$A_{\bar{0}}$ ( resp.
$A_{\bar{1}}$). Let $\hat{A}$ be the
\emph{Clifford affinization} of $A$, which is
the Lie superalgebra $\hat{A} = A \otimes
\CZ [t, t^{-1}] + \CZ K$ with the commutation
relations
$$
[ a t^m , b t^n ] = \langle a , b \rangle \delta_{m, -n-1} K \ ,
\ \ [ K , \hat{A} ] = 0 \ .
$$
We take the filtration $\hat{A}_j=\CC K+\sum_{i \geq j} At^i$ for
$j \leq 0$,     $\hat{A}_{(j)} = \sum_{i \geq j}
A t^i$ for $j>0$, and let $k = 1$.
For $\Phi \in A$, let $\Phi (z) =
\sum_{n
\in \ZZ} (\Phi t^n) z^{-n-1}$. Then $\{ \Phi (z)
\}_{ \Phi
\in A }$, called a  collection of
\emph{neutral free superfermions}, which consists of
pairwise local fields with $\lambda$-bracket
$$
[ \Phi_\lambda \Psi ] = \langle \Phi , \Psi
\rangle 1
\ ,
\ \,
\ \ \Phi , \Psi \in A \ .
$$
Let $\{ \Phi_i \}$ and $\{ \Phi^i \}$ be a pair of
dual  bases of $A$, i.e.,~$\langle
\Phi_i ,\Phi^j
\rangle = \delta_{i,j}$, and define
\be
L = \frac{1}{2} \sum_{i} : (\partial \Phi^i )
\Phi_i :  \ .
\ele(fsfL)
Then $L$ is an energy-momentum field with
central charge $c = - \frac{1}{2} {\rm sdim} \
A$.
Furthermore, the neutral free
superfermions
$\Phi (z)$ are all
primary (with respective to this $L$) of
conformal weight
$\frac{1}{2}$. The vertex algebra
$F(A)$ strongly generated by these superfermions,
with the above energy-momentum field $L$, is
called the \emph{vertex algebra of neutral free
superfermions}.  Via the state-field correspondence, $F(A)$ is identified with the space $U_1(\hat{A})/U_1 (\hat{A}) \hat{A}_{(0)}$, and all fields of $F(A)$ act on this space from the left.

{\bf Example 1.3} (charged free superfermions).
Let $A_{\rm ch}$ be a
finite-dimensional superspace with a
non-degenerate skew-supersymmetric bilinear form
$\langle . \ , \ . \rangle$, and suppose that
$A_{\rm ch} = A_+ \oplus
A_-$, where both $A_\pm$
are isotropic subspaces of $A_{\rm
ch}$. We have the Clifford affinization
$\hat{A}_{\rm ch}$ with the filtration $(\hat{A}_{\rm ch})_{(j)}
\ (j \in \ZZ_+)$, $k=1$, and  the fields $\varphi (z),
\varphi^*(z)$ for $\varphi
\in A_+ , \varphi^* \in
A_-$, as in Example~1.2. We define
the {\it charges} of the fields by
\be
{\rm charge} \ \varphi (z) = - \ {\rm charge} \
\varphi^* (z) = 1 \ .
\ele(chdef)
Let $\{ \varphi_i \}$ ( resp. $\{ \varphi^*_i
\}$) be a  basis of $A_+$ (resp.
$A_-$)  such that $\langle \varphi_i ,
\varphi^*_j \rangle = \delta_{i, j}$. The set of
pairwise local fields $\{ \varphi_i (z) \} \cup \{
\varphi^*_i (z)
\}$ is called a collection of {\it charged free
superfermions}. In this case, we can define a
family of energy-momentum fields parametrized by
$\vec{m} = (m_i)_{i} , m_i \in \CZ$:
$$
L^{\vec{m}}  = - \sum_{i} m_i : \varphi^*_i
\partial \varphi_i: + \sum_{i} ( 1 - m_i) :
(\partial \varphi^*_i) \varphi_i : \ .
$$
The central charge of $L^{\vec{m}} (z)$ is equal
to
$$
- \sum_{i} (-1)^{p(\varphi^i)} ( 12 m_i^2 - 12
m_i + 2 )
\ .
$$
Furthermore, the fields $\varphi^*_i (z)$ and $\varphi_i
(z)$ are  primary ( with respect to
$L^{\vec{m}}$ ) of conformal weights $m_i$ and
$1- m_i$ respectively. The vertex algebra $F (
A_{\rm ch})$ with one of the
energy-momentum fields $L^{\vec{m}}$ is called
the {\it vertex algebra of charged free
superfermions}. The relations (\req(chdef)) give
rise to the {\it charge decomposition} of $F (
A_{\rm ch})$:
\be
F ( A_{\rm ch}) = \bigoplus_{m \in \ZZ} F_m ( A_{\rm ch}) \ .
\ele(cdec)

{\bf Example 1.4} (currents and the Sugawara
construction). Let $\goth g$
be a simple finite-dimensional Lie superalgebra
with an even non-degenerate supersymmetric invariant
bilinear form $( . | . )$. Let $\hat{{\goth g}}$ be the
Kac-Moody affinization of $\goth g$,
i.e.,~$\hat{{\goth g}} = \goth g \otimes \CZ
[ t, t^{-1} ]  \oplus \CZ K \oplus \CZ D $ with
the commutation relations:
$$
[at^m , b t^n ] =[a, b]t^{m+n} + m
\delta_{m, -n}(a|b)K \ ,  \ \ [ D , a t^m ] = m a
t^m \ , \ \
\ [K,
\hat{\goth{g}} ]=0
\ .
$$
The filtration in this situation is defined as in Example~1.2,
 and we fix $k \in \CC$.

For an element $a \in \goth g$, one
associates the {\it current} field $a(z) =
\sum_{n \in
\ZZ} ( a t^n ) z^{-n-1}$. The collection $\{ a(z)
\}_{a \in \goth g}$ consists of
pairwise local fields with the following
$\lambda$-bracket,
$$
[ a_\lambda b ] = [ a , b ] + \lambda (a | b ) k
\ , \ \ a , b \in \goth g \ .
$$
The vertex algebra $V_k(\goth g)$ strongly generated by the current fields $a(z)$
is called the {\it universal affine vertex algebra}. Via the
state-field correspondence, $V_k (\goth g)$ is identified with the
space $U_k (\hat{\goth g})/U_k (\hat{\goth g}) \hat{\goth g}_{(0)}$ and
all fields of $V_k (\goth g)$ act on this space from the left.

Let $\{ a_i \}$ and $\{ a^i \}$ be a pair of dual bases of
$\goth g$: $( a_i | a^j ) = \delta_{i,
j}$. Then $\Omega = \sum_i(-1)^{p(a_i)} a_i a^i$ is the Casimir
operator of $\goth g$, and it lies in
the center of $U ( \goth g )$.
The one-half of the eigenvalue of $\Omega$ in the
adjoint representation, denoted by
$h^\vee$, is called the {\it dual Coxeter
number} of $\goth g$ (it depends on
the normalization of $( . | . )$).

Recall the following relation between the Killing form and the
form $(.|.)$ \cite{KW3}:
\begin{equation}
  \label{eq:1.7}
  {\rm str}_{\goth g} ({\rm ad} \,a) ({\rm ad}\, b) =2h^{\vee} (a|b)\, ,
  \quad a,b \in \goth g \, .
\end{equation}
(Since the LHS is the Killing form, it is equal to $\gamma (a|b)$
for some $\gamma$.  Hence ${\rm str}_{\goth g} \Omega =\gamma \,
{\rm sdim}\,  \goth g$.  Since $\Omega =2h^{\vee} I_{\goth g}$, we
conclude that $\gamma =2h^{\vee}$ provided that ${\rm sdim}\,  \goth
g \neq 0$.  Hence (\ref{eq:1.7}) holds for all exceptional Lie
superalgebras and also for all the series $s \ell (m|n)$,
etc. apart for the values $(m,n)$ on a hyperplane.  Hence
(\ref{eq:1.7}) holds for all values $(m,n)$.)

Assuming $k +
h^\vee \neq 0$, introduce the so called
{\it Sugawara construction}:
$$
L(z) = \frac{1}{2(k + h^\vee)} \sum_i (-1)^{p(a_i)} : a_i (z) a^i
(z) \, .
$$
This is an energy momentum field with the central
charge
\be
c ( k ) = \frac{k \ {\rm sdim}
\goth g }{k + h^\vee} \ .
\ele(cSug)
All currents  are primary with respect
to $L$ of conformal weight 1. We shall also
use the following well known modification of the
Sugawara construction. For a given $a \in
\goth g_{\bar{0}}$, let
$$
L^{(a)} = L + \partial a \ \ .
$$
This is again an energy momentum field, and its
central charge becomes
\be
c ( k , a ) = c ( k ) - 12 k ( a | a ) \ .
\ele(cka)
With respect to $L^{(a)}$, the currents are not
 primary anymore:
\be
[ {L^{(a)}}_\lambda b ] = \partial b + \lambda
( b - [ a , b ] ) - \lambda^2 k ( a | b ) \ .
\ele(Lab)
However, one has
\be
[ {L^{(a)}}_\lambda b ] = \big( \partial + (1-m)
\lambda \big)  b \ , \ \ \ {\rm if} \ \ [ a , b ]
= m b \ , \ m \neq 0 \ ,
\ele(Labp)
since in this case $(a | b ) = 0 $.

\section{The Quantum Reduction}
\setcounter{equation}{0}
\subsection{The complex ${\cal C}
(\goth g, x,  f, k)$ and the associated
vertex algebra $W_k(\goth g, x, f)$}

Here we describe a general construction of a vertex algebra via a
differential complex, associated to a simple finite-dimensional
Lie superalgebra and some additional data, by a quantum reduction
procedure, generalizing that of \cite{FF1}, \cite{FF2}, \cite{FKW}, \cite{BT}.

Let $\goth g$ be a simple finite-dimensional Lie
superalgebra with a non-degenerate even supersymmetric invariant
bilinear form $( . | . )$.  Fix a pair $x$ and $f$ of even
elements of $\goth g$ satisfying the following
properties:

\Alist
\begin{enumerate}
\item 
$\ad x$ is diagonizable with half-integer eigenvalues, i.e.,~we
have the following eigenspace decomposition with respect to $\ad
x$:
\begin{equation}
  \label{eq:2.1}
  \goth g = \oplus_{j \in \frac{1}{2} \ZZ}
    \goth g_j \, .
\end{equation}

\item 
$
f \in \goth g_{-1} \, , \hbox{ i.e.,~} [x,f]=-f \, .
$
\end{enumerate}
It follows
that $f$ is a nilpotent element of $\goth g$. We shall
also assume

\begin{enumerate}
\setcounter{enumi}{2}
\item 
$\ad f: \goth g_{\frac{1}{2}} \to {\goth g}_{-\frac{1}{2}}$ is a vector space isomorphism.
\end{enumerate}

The element $f$ defines a skew-supersymmetric even
bilinear form on  $\goth g_{\frac{1}{2}} $ by the
formula:
\be
\label{eq:2.2}
\langle a , b \rangle = ( f | [ a , b ] ) \ .
\ee
It follows from (A3) that
this form is
non-degenerate, since  
$\langle a , b \rangle = ( [ f , a ] | b )$
 and  $( . | . )$ gives a
non-degenerate pairing between ${\goth g}_{- \frac{1}{2}}$ and
 ${\goth g}_{\frac{1}{2}}$.
Denote by
$A_{\rm ne}$ the vector superspace $
\goth g_{\frac{1}{2}} $ with the
non-degenerate supersymmetric bilinear
form $\langle \cdot \ , \ \cdot
\rangle$.

Furthermore, let
\be
\goth g_+ = \bigoplus_{j >0} \goth g_j \ , \,
\goth g_- = \bigoplus_{j<0} \goth g_j \, ,
\ele(g+)
and let
$$
A = \sqcap \goth g_+ \ ,
\ \ A^* = \sqcap {\goth g}_+^* \ , \ \
A_{\rm ch} = A \oplus
A^* \ ,
$$
where $\sqcap$ stands the reversing the parity
of  a vector superspace. Let $\langle \cdot \ , \ \cdot
\rangle$ be the skew-supersymmetric bilinear
form on $A_{\rm ch}$ defined by
$$
\langle A , A \rangle = \langle A^* , A^* \rangle = 0 \ , \ \ \langle a , b^* \rangle = b^* (a) \ \ \ {\rm for} \ a \in A , \ b^* \in A^* \ .
$$
Define gradations of $A,
A^*$ by (\req(g+)):
$$
A = \bigoplus_{j>0}
A_j \ , \ \ \ \ A^* =
\bigoplus_{j>0}
A_j^* \ .
$$
Finally, fix a complex number $k$ such that $k+h\sucheck \,\neq 0$,
where $h\sucheck$ is the dual Coxeter number of $\goth g$.

We shall associate to the data $({\goth g}, x,f,k)$ a
differential vertex algebra $(\C({\goth g},x,f,k), d_0)$
(by this we mean that $\C$ is a vertex algebra and $d_0$ is
an odd derivation of all $n$-th products of $\cal C$, and  $d^2_0 =0$).

Let $\hat{\goth g}, \hat{A}_{\rm ne}$ and $\hat{A}_{\rm ch}$ be
the Kac--Moody and Clifford affinizations
corresponding to $\goth g,
A_{\rm ne}$ and $A_{\rm
ch}$ respectively (see Examples 1.4, 1.2 and
1.3). Let $U_k = U_k ( \hat{\goth g} ) \otimes U_1 (
\hat{A}_{\rm ch} ) \otimes U_1 ( \hat{A}_{\rm ne})$, and
let $U_k^{\rm com}$ be the completion of $U_k$ as
defined in Section 1. Consider the corresponding
vertex algebras $V_k ( \goth g )$, $ F
(A_{\rm ch})$ and $F (
A_{\rm ne} )$, generated by the
currents (based on $\goth g$),
charged free fermions (based on
$A_{\rm ch}$), and neutral free
fermions ( based on $
A_{\rm ne}$) respectively. Consider the
vertex algebras
$$
F ( \goth g,x,  f ) =  F (
A_{\rm ch})
\otimes F ( A_{\rm ne} ) \ , \ \ \
{\cal C} ( \goth g,x, f , k ) = V_k (
\goth g )
\otimes F ( \goth g,x, f ) \ .
$$
By letting ${\rm charge} (V_k ({\goth
g})) =  {\rm charge} (F ( A_{\rm ne}
)) = 0$, and using (\req(cdec)), one has the
induced charge decompositions of $F (
\goth g, f )$ and
$ {\cal C} ( \goth g, f , k )$:
$$
F ( \goth g,x, f )= \bigoplus_{m \in
\ZZ} F_m \ , \ \ \ {\cal C} ( \goth g,x,
f , k ) = \bigoplus_{m \in \ZZ} {\cal C}_m \ .
$$

Next, we define a differential
on
${\cal C} ( \goth g, x,f , k )$, which
makes it a homology complex.  For this purpose,
choose a basis $\{u_i \}_{i \in S^\prime}$ of
$\goth g_{\frac{1}{2}}$, and extend
it a basis $\{u_i \}_{i \in S}$ of
$\goth g_+$ compatible with the
gradation (\req(g+)). Furthermore, extend the latter basis to a
basis $\{ u_i\}_{i \in \tilde{S}}$ of $\goth g$, compatible
with this gradation, and define the
structure constants $c^\ell_{ij}$ by: $ [u_i , u_j] = \sum_\ell c^\ell_{i
j} u_\ell$.  Denote by $\{ u^i \}_{i \in S'}$ the dual basis of
$\goth g_{\tfrac{1}{2}}$ with respect to the form $\langle \, ,
\, \rangle$, i.e.,~$\langle u_i , u^j \rangle =\delta_{ij}$.

Denote by $\{ \varphi_i \}_{i \in S}, \{
\varphi_i^\ast \}_{i \in S}$ the corresponding
bases of $A$ and $A^*$, and by
$\{\Phi_i
\}_{i \in S^\prime}$ the corresponding basis of
$A_{\rm ne}$. The fields
$\varphi_i(z), \varphi_i^\ast (z) \ (i \in S)$
and $\Phi_i (z) \ (i \in S^\prime)$ are called
\emph{ghosts}. Introduce the
following field of the vertex algebra ${\cal C} (
\goth g,x, f , k)$:
$$
\begin{array}{ll}
d (z) =& \sum_{i \in S } (-1)^{p(u_i)}
u_i(z)
\otimes \varphi^\ast_i (z) \otimes 1 -
\frac{1}{2} \sum_{i, j , \ell \in S } (-1)^{p(u_i)
p(u_\ell)} c^\ell_{i j }  \otimes \varphi_\ell
(z)
\varphi^\ast_i (z) \varphi^\ast_j
(z) \otimes 1 \\
 &+ \sum_{i \in S} (f|u_i)  \otimes
\varphi^\ast_i(z) \otimes 1 + \sum_{i \in
S^\prime} 1 \otimes
\varphi^\ast_i (z) \otimes \Phi_i
(z) \, .
\end{array}
$$
For simplicity of notation, we shall
omit the tensor sign
$\otimes$ in the expression of fields. Note that in the
second term of the expression of $d(z)$, one has
$$
\varphi_\ell (z) \varphi^\ast_i (z) \varphi^\ast_j (z) = : \varphi_\ell
(z) \varphi^\ast_i (z) \varphi^\ast_j
(z): \ \ \ {\rm if } \ c^{\ell}_{ij} \neq 0 \ ,
$$
hence $d(z)$ is a vertex algebra field. Also, it is easy to see that
$d(z)$ is an odd field independent of the choice
of the basis.
By the right non-commutative Wick
formula one has the following
$\lambda$-brackets of $d(z)$ and the currents
$u_j(z) \,\, (j \in \tilde{S})$, and the ghosts
$ \varphi_j (z),  \varphi_j^*(z)
\, (j \in S)$ and $ \Phi_j(z) \,\, (j \in S')$:
\bea(lll)
&[ d_\lambda u_j ]  = &
\sum_{\substack{i \in S \\ \ell \in \tilde{S}} }
(-1)^{p(u_j)+p(u_\ell)p(u_i)}  c^\ell_{ij} u_\ell
\varphi^\ast_i + (\partial +\lambda) k \sum_{i \in S
}(u_j | u_i ) \varphi^\ast_i \ ; \\[1ex]
&[d_\lambda \varphi_j]  = &  u_j +   (f|u_j)   +
\sum_{i,  \ell \in S } (-1)^{p(u_\ell)} c^\ell_{j i }
\varphi_\ell  \varphi_i^*
 + \sum_{i \in S^\prime}
(-1)^{p(u_i)} \delta_{i, j}  \Phi_i \ ; \\[1ex]
& [d_\lambda \varphi^\ast_j] = & -
\frac{1}{2} \sum_{ i , s \in S } (-1)^{p(u_i)
p(u_j)} c^j_{i s }   \varphi^\ast_i
\varphi^\ast_s \ ; \\[1ex]
& [d_\lambda \Phi_{j}] = &
\sum_{i \in S^\prime} ( f | [ u_i , u_{j} ] )
\varphi^\ast_i \, , [d_{\lambda}\Phi^j]=\varphi^*_j \ .
\elea(df)
\begin{theorem} \label{thm:dd}
One has:  $[ d ({z})_\lambda d ({z})] = 0$.
\end{theorem}
{\it Proof.} We express the field $d(z)$ as
$$
d(z) = d (z)^{\rm st}  + d
(z)^{(III)} + d (z)^{(IV)} \ , \ \ \ d (z)^{\rm
st} := d (z)^{(I)} + d (z)^{(II)} \ ,
$$
where
$$
\begin{array}{ll}
d^{(I)} = \sum_{i \in S } (-1)^{p(u_i)}
u_i \varphi^\ast_i \ , & d^{(II)} =
\frac{-1}{2} \sum_{i, j , \ell \in S } (-1)^{p(u_i)
p(u_l)} c^l_{i j }  \varphi_\ell
\varphi^\ast_i  \varphi^\ast_j \ , \\
d^{(III)} = \sum_{i \in S} (f|u_i)
\varphi^\ast_i \ ,  & d^{(IV)}= \sum_{i
\in S^\prime}  \varphi^\ast_i  \Phi_i \ .
\end{array}
$$
Then
$$
[d_\lambda d] = [d^{\rm st}_\lambda d^{\rm st}] +
[d^{(II)}_\lambda d^{(III)}] + [d^{(III)}_\lambda
d^{(II)}] + [{d^{(II)}}_\lambda \, d^{(IV)}] +
[{d^{(IV)}}_\lambda\,  d^{(II)}] + [{d^{(IV)}}_\lambda \,
d^{(IV)}] \ .
$$
It is well known that $[d^{\rm st}_\lambda d^{\rm
st}]=0 $, which follows from $[{d^{(II)}}_\lambda \,
d^{(II)}] = 0$ by the Jacobi identity.  In the
expression of
$d^{(II)}$,  one has $\ell \not\in S^\prime$
whenever $c_{ij}^\ell \neq 0$, hence
$[\varphi_{\ell \lambda} \varphi_k^\ast] = 0$ for $k \in
S^\prime$. This implies  $[{d^{(II)}}_\lambda\,
d^{(IV)}] = [{d^{(IV)}}_\lambda \, d^{(II)}] = 0$.
Hence
$$
[d_\lambda d] = [{d^{(II)}}_\lambda \, d^{(III)}] +
[{d^{(III)}}_\lambda\,  d^{(II)}] +  [{d^{(IV)}}_\lambda \,
d^{(IV)}]  \ .
$$
Note that  $p(u_\ell)=
p(u_i)+ p(u_j)$ whenever $c^\ell_{i j} \neq 0$. We
have
$$
\begin{array}{lll}
&[{d^{(IV)}}_\lambda\,  d^{(IV)}] &= \sum_{i,j \in
S^\prime} [{\varphi^\ast_i  \Phi_i}_\lambda
\varphi^\ast_j \Phi_j ] = \sum_{i,j \in
S^\prime} (-1)^{p(u_i)(p(u_j)+1)}
\varphi^\ast_i \varphi^\ast_j ( f | [u_i , u_j ])
\\[1ex]
 && = \sum_{i,j \in
S^\prime , \ell \in S } (-1)^{p(u_i)(p(u_j)+1)}
c_{i j}^\ell ( f | u_\ell) \varphi^\ast_i \varphi^\ast_j
\ ; \\[1ex]
&[{d^{(II)}}_\lambda\,  d^{(III)}] &=
\frac{-1}{2} \sum_{i, j , \ell \in S } (-1)^{p(u_i)
p(u_\ell)} c^\ell_{i j } (f | u_\ell) [ {  \varphi_\ell
\varphi^\ast_i  \varphi^\ast_j }_\lambda
\varphi_\ell^\ast ] \\[1ex]
&&=
\frac{-1}{2} \sum_{i, j , \ell \in S } (-1)^{p(u_i)
( p(u_i)+p(u_j))} c^\ell_{i j } ( f | u_\ell)
\varphi^\ast_i  \varphi^\ast_j \\[1ex]
&& =
\frac{-1}{2} \sum_{i, j \in S^\prime , \ell \in S }
(-1)^{p(u_i) ( p(u_j)+1)} c^\ell_{i j } (f| u_\ell)
\varphi^\ast_i  \varphi^\ast_j \ \ ({\rm since} \
i, j \in S^\prime \ {\rm if} \ (f|[u_i, u_j]) \neq
0 ) \ .
\end{array}
$$
Therefore $[{d^{(II)}}_\lambda \, d^{(III)}] =
[{d^{(III)}}_\lambda \, d^{(II)}]= \frac{-1}{2}
[{d^{(IV)}}_\lambda \, d^{(IV)}]$, hence $[d_\lambda
d]=0$.
 $\Box$ \par \vspace{.2in}
Let $d_0 = {\rm Res}_z d (z)$. Note that $d_0$ is
an odd element of $U_k^{\rm com}$, and that
$[d_0
, {\cal C}_m] \subset {\cal C}_{m-1}
$.
Theorem \ref{thm:dd} implies that $[d(z),d(w)]=0$, hence $[d_0, d_0]
= 2 d_0^2 = 0$. Thus $( {\cal C}(
\goth g, x, f , k), d_0 )$ is a
homology complex. We denote the $0$-th homology
of this complex by $
W_k ( \goth g, x,f ) $.
Since ${\cal C}_0$ is a vertex subalgebra of ${\cal C}(
\goth g, f , k)$, and since $d_0$ is a derivation of all of its $n$\st{th} products, we conclude that $W_k (
\goth g,x, f )$ is a vertex algebra. This
vertex algebra is called the {\it quantum
reduction} for the quadruple $(\goth g,x, f ,
k)$.

The most interesting pair $x,f$ satisfying properties (A1), (A2),
(A3) comes from an $s\ell_2$-triple $\{ e,x,f \}$, where
$[x,e]=e$, $[x,f]=-f$, $[e,f]=x$. The validity of these
properties is immediate by the $s \ell_2$-representation theory.
Since a nilpotent even element $f$ determines uniquely (up to conjugation) the
element $x$ of an $s\ell_2$-triple (by a theorem of Dynkin), we
shall use in this case the notation $W_k (\goth g,f)$ for the
quantum reduction.

The vertex algebra $W_k (\goth g,f)$ is a generalization of the quantum
Drinfeld--Sokolov reduction, studied in \cite{FF1}, \cite{FF2},
\cite{FKW}  and many other papers, when ${\goth g}$ is a simple Lie algebra
and $f$ is the principal nilpotent element. The
case studied in
\cite{B}  is when ${\goth g} = {\goth sl}_3$ and
$f$ is a non-principal nilpotent element. Our
construction is a development of the
generalizations proposed in \cite{FKW} and in
\cite{BT}.  \par
\vspace{.1in}
\noindent
{\bf Remark~2.1.}~~(a)~~The assumption (A3) is not used in the proof of Theorem~2.1.
However, this condition is essential for the construction of the
energy-momentum field $L(z)$ in Section~2.2.

(b)~~One can take for $x$ a diagonalizable derivation of $\goth g$.

(c)~~ Let ${\goth n}$ be an  ${\rm ad }  x$-invariant subalgebra of
$\goth g_+$. The above construction when applied to ${\goth n}$
in place of $\goth g_+$ produces a complex $ ({\cal C}( \goth g,
{\goth n}, x,f , k), d_{{\goth n}} )$. The corresponding vertex
algebra $W_k ( \goth g, \goth n,x, f )$ is naturally a
subalgebra of $W_k ({\goth g}, x,f )$.

\subsection{The energy-momentum field of
$ W_k ( \goth g, x, f)$}
Denote by $L^{\goth g}(z)$  the Sugawara energy
momentum field of $\hat{\goth g}$
(see Example 1.4), by $L^{\rm ne}$ the energy
momentum field for $F ( A_{\rm ne} )$
(see Example 1.2), and by $L^{\rm ch}$
the energy momentum field $L^{\vec{m}} $ for $F (
A_{\rm ch} )$ (see Example 1.3) with
$m_i$'s defined by
$$
[ x , u_i ] = m_i u_i \ .
$$
Let
\be
L (z) = L^{\goth g} (z) + \partial_z x (z) +
L^{\rm ch} (z) +
L^{\rm ne} (z) \ .
\ele(LCom)
The discussion in Section 1 immediately implies
the following result.
\begin{theorem} \label{thm:Cgfk}
{\rm (a)} The field $L(z)$ is the energy-momentum
field  for the vertex algebra ${\cal
C}(\goth g,x,  f , k)$, and its central
charge equals to
\be
c (\goth g ,x,f,k)   =\frac{k \ {\rm sdim} \goth g}{k + h^\vee
} - 12 k (x | x ) -
\sum_{i
\in S} (-1)^{p(u_i)} ( 12 m_i^2 - 12 m_i + 2 ) -
\frac{1}{2} {\rm sdim} \
\goth g_{ \frac{1}{2}} \ .
\ele(ckx)

{\rm (b)} With respect to $L(z)$, the fields
$\varphi_i (z), \varphi_i^* (z) \ (i \in S)$, are
primary of conformal weights $1-m_i, m_i$
respectively, and the fields $\Phi_i (z) \ (i \in
S^\prime)$, are primary of conformal weight
$\frac{1}{2}$. The fields $u (z)$ for $u \in
\goth g_j$ have conformal weight
$1-j$, and are primary unless $j = 0 $ and $(x |
u ) \neq 0$.
\end{theorem}
$\Box$ \par \vspace{.2in}

\textbf{Remark 2.2.}  In the same way as in \cite{FKW}, formula (2.6) can
be rewritten as follows (see also \cite{BT}):
\begin{displaymath}
  c  (\goth g ,x,f,k)   ={\rm sdim}\, \goth g_0-\frac{1}{2} \, {\rm sdim} \,
  \goth g_{1/2}-12 | \frac{\rho}{(k+h\sucheck\,)^{1/2}}
  -x(k+h\sucheck\,)^{1/2}|^2 \, .
\end{displaymath}

The next theorem says that the field $L (z)$ defined by (\req(LCom))
is the energy-momentum field for the
vertex algebra
$W_k ( \goth g,x,  f )$.
\begin{theorem} \label{thm:dL} We have $[ d_0 ,
L(z)] = 0 $.
\end{theorem}
{\it Proof.} We compute the $\lambda$-bracket
$[ L_\lambda d]$.
Using Theorem \ref{thm:Cgfk} (b) and the Wick
formula (the "non-commutative" terms vanish
everywhere) , we have ( recall that $u_i \in
\goth g_+$):
$$
\begin{array}{lll}
&[ L_\lambda (u_i \varphi^\ast_i) ] &  = \partial ( u_i \varphi^\ast_i ) +  \lambda u_i \varphi^\ast_i \ , \\[1ex]
&[ L_\lambda (\varphi^\ast_i  \varphi^\ast_j) ] &= \partial ( \varphi^\ast_i  \varphi^\ast_j) +  (m_i+m_j) \lambda \varphi^\ast_i \varphi^\ast_j \ , \\[1ex]
&[ L_\lambda (\varphi_\ell \varphi^\ast_i  \varphi^\ast_j) ]&= \partial (\varphi_\ell \varphi^\ast_i  \varphi^\ast_j) + (1-m_\ell+ m_i+m_j) \lambda \varphi_\ell  \varphi^\ast_i \varphi^\ast_j \ , \\[1ex]
&[ L_\lambda (\varphi^\ast_i \Phi_i) ]&  =
\partial (\varphi^\ast_i  \Phi_i) +
(\frac{1}{2}+m_i) \lambda \varphi^\ast_i \Phi_i \
,
\end{array}
$$
therefore
$$
[ L_\lambda d ] = ( \partial + \lambda ) d + \lambda ( \sum_{i
\in S} (m_i - 1) (f|u_i) \varphi^\ast_i  +
\sum_{i \in S^\prime} (m_i - \frac{1}{2})
\varphi^\ast_i \Phi_i  ) \ .
$$
Since $(f | u_i ) = 0$
unless $m_i =1$, and $m_i = \tfrac{1}{2}$ if $i \in
S^\prime$, we have
$ [ L_\lambda d ] = (
\partial + \lambda ) d$. Hence, by
skew-commutativity,
$ [ d_\lambda L ] = \lambda d$, and therefore, by
Corollary \ref{cor:a0b0}, $[ d_0 , L(z)] = 0 $.
$\Box$ \par \vspace{.2in}

\subsection{The quasiclassical limit }
 Here we briefly discuss the
standard construction of the quasiclassical limit
for the complex ${\cal C} ( \goth g,x,
f, k )$.

Denote by $A_\hbar$ the space of the Lie
superalgebra $A$ with the new bracket.
$$
[ a , b ]_\hbar = \hbar [ a , b ] \ , \ \ \ \ a , b \in
A \ .
$$
Then $U_k ( A_\hbar )$ is the quotient of the
tensor algebra over the vector space  $A$
by the ideal generated by the elements $(K - k)$
and $a \otimes b - (-1)^{p(a) p(b)} b \otimes a -
\hbar [ a , b ] ( a , b \in A)$. Hence the limit
of
$U_k ( A_\hbar )$ as $\hbar \rightarrow 0$ is
 $S_k (A)$, the symmetric superalgebra over $A$
quotiented by the ideal $(K - k)$,
with the Poisson bracket:
$$
\{ u , v \} = \lim_{\hbar \rightarrow 0}
\frac{1}{\hbar} [ u , v ]_\hbar \ , \ \ \ u , v \in S_k
(A) \ .
$$
In the same way as in Section 1, we construct the
Poisson superalgebra $S_k
(A)^{\rm com} \supset S_k (A)$.

As in Section 1, we consider $S_k (A)^{\rm
com}$-valued fields, and define their $n$-th
product for $n \in \ZZ_+$, by $a(z)_{(n)}b(z) =
{\rm Res}_x ( x-z)^n \{ a (x), b(z) \}$, and let
$\{ a_\lambda b \} = \sum_{n \in \ZZ_+}
\frac{\lambda^n }{n!} a_{(n)} b$. Since the
product in $S_k (A)^{\rm com}$ is
(super)commutative, the normal ordered product
becomes the usual product. Then Proposition
\ref{prop:lambda} holds for $\{ a_\lambda b \}$,
except  that ``non-commutative'' Wick product
formula turns into the  Leibniz rule:
$$
\{a_\lambda  bc  \} = \{ a_\lambda b \}c +
(-1)^{p(a) p(b)} b \{ a_\lambda c \} \ .
$$
Proposition \ref{prop:lamloc} $(a)$ and $(b)$
hold as well, while $(c)$ turns into the
supercommutativity of the product.
 The vertex algebra of free superfermions of
Examples 1.1 and 1.2 in the quasiclassical limit
turns into the Poisson vertex algebra generated
by the fields $\{ a ( z ) \}_{a \in
A}$ with the $\lambda$-bracket
$$
\{ a_\lambda b \} = \langle a , b \rangle 1 .
$$
All formulas of Example 1.2---1.4
hold in  the limit, except that the Virasoro
central charge becomes 0 for Examples 1.2, 1.3
and in the formula (\req(cSug)) in Example 1.4,
hence (\req(cka)) becomes $-12 k (a|a)$. Thus,
the central charge of $L(z)$ in the limit becomes
$-12k(x|x)$ (cf (\req(ckx))).

The method of quantum reduction turns the Poisson
structure of the quasiclassical limit into the
generalized Drinfeld-Sokolov reduction. The
complex ${\cal C}(\goth g,x, f , k )$
turns into the tensor product of the corresponding
Poisson vertex algebras, and the differential $d$
is given by the same formula, (except that the
commutator with $d$ is replaced by the Poisson
bracket with $d$). Finally, the energy-momentum
field is given by the same formula, but the
central charge is $- 12 k ( x | x )$.

\subsection{The basic conjecture on the
structure of
$ W_k ( \goth g,x, f)$}
Let $\goth g^f$ be the centralizer of
$f$ in $\goth g$. The gradation
(2.1) induces a $\frac{1}{2}\ZZ$-gradation
\be
\goth g^f = \displaystyle{\bigoplus_j} {\goth g}^f_j \ .
\ele(gf)

For a good description of the vertex algebra $W_k
(\goth g,x,f)$ the following additional condition is
apparently necessary:

\Alist
\begin{enumerate}
\setcounter{enumi}{3}
\item  
The operator $\ad f$ maps $\goth g_j$ to $\goth g_{j-1}$
injectively for $j \geq 1$ and surjectively for $j\leq
0$.
\end{enumerate}

By the representation theory of $s\ell_2$, this condition
holds if the pair $x,f$ can be embedded in an $s\ell_2$-triple
(but there are many more examples).

We shall call a pair $(x,f)$ satisfying conditions (A1)--(A4) to
be a \emph{good pair}, and the one coming from an
$s\ell_2$-triple a \emph{Dynkin pair}.  The corresponding
$\frac{1}{2}\ZZ$-gradations are called \emph{good} and
\emph{Dynkin} gradations, respectively.  Note that these
gradations uniquely determine $x$ (by definition) and also
determine $f$ up to conjugation by $G_0=\exp (\goth g_{0,
\bar{0}})$ (preserving the gradation), since $[\goth g_{0,
\bar{0}},f] = \goth g_{-1, \bar{0}}$ and therefore $f$ lies in the
open orbit of $G_0$.

{\bf Conjecture 2.1} Suppose that conditions (A1)--(A4) hold.
Then for each $a \in
\goth g^f_{-j} \ ( j \geq 0)$ there
exists a field $F_a (z)$ of the vertex
algebra ${\cal C} ( \goth g,x, f , k )$,
such that the following properties hold:

(i) $[ d_0, F_a (z) ] = 0$ ,

(ii) $F_a (z)$ has conformal weight $1+j$ with
respect to $L (z)$,

(iii) $F_a (z) - a(z)$ is a linear combination of
normally ordered products of
the fields $b(z)$, where $b \in
\goth g_s$ with $s > -j$, the
ghosts $\varphi_i (z), \varphi_i^\ast (z), \Phi_i
(z)$, and their derivatives. \par \noindent
Furthermore, the images of the fields $F_{a_i}
(z)$ in $W_k ( \goth g,x,  f  )$, where
$\{ a_i \}$ is a basis of $\goth g^f$
compatible with the gradation (\req(gf)),
strongly generate the vertex algebra $W_k (
\goth g, x, f  )$.

Given $v \in  \goth g $, introduce
the fields (we assume here condition (A3)):
\begin{eqnarray*}
v^{\rm ch} (z) &= - \sum_{i, j \in S}
(-1)^{p(\varphi_i)} c_{i j} (v): \varphi_i (z)
\varphi_j^\ast (z) : \ , \\
v^{\rm ne} (z) & = - \frac{1}{2}
\sum_{i, j\in S^\prime}(-1)^{p (\Phi_i)}  c_{i j} (v): \Phi_i (z) \Phi^j (z):
 \ ,
\end{eqnarray*}
where the $c_{i j} (v)$ are defined by $[v , u_j ] =
\sum_{i} c_{i j}(v) u_i$ and , as before,
$\langle \Phi_i, \Phi^j \rangle = \delta_{i j}$.
  Note that $v^{\rm ne}(z)=0$ unless $v \in \goth g_0$, and that all
  pairs of distinct fields from $\{ v, v^{\rm ch}, v^{\rm ne} \}$ have
  zero $\lambda$-brackets.  Let
$$
J^{\{ v\}} (z) = v(z) + v^{\rm ch}(z) + v^{\rm ne} (z) \, , \,
J^{(v)} (z) = v (z) + v^{\rm ch} (z) \, .
$$

The calculations with $v^{\rm ch}$ and $v^{\rm ne}$ will use the
following lemma.

\begin{lemma}
  \label{lem:2.1}
\alphaparenlist
  \begin{enumerate}
  \item 
    Let $v \in \goth g_0$.  Then
    \begin{eqnarray*}
      [{v^{\rm ch}}_{\lambda} \varphi_k]
        &=& (-1)^{p(v)} \sum_{i \in S} c_{ik}(v)
            \varphi_i \, , \\[1ex]
     [{v^{\rm ch}}_{\lambda} \varphi^*_k]
        &=& -(-1)^{p(v)p(\varphi^*_k)} \sum_{j \in S}
            c_{kj} (v) \varphi^*_j \,  .
    \end{eqnarray*}
\item 
  Let $v \in \goth g^f_0$.  Then
  \begin{displaymath}
    [{v^{\rm ne}}_{\lambda} \Phi_k] =(-1)^{p(v)}
       \sum_{i \in S'} c_{ik} (v) \Phi_i \, .
  \end{displaymath}

  \end{enumerate}
\end{lemma}

{\it Proof.}  The proof of (a) is straightforward, using the Wick
formula and the observation that
\begin{equation}
  \label{eq:2.8}
  p(u_i) + p(u_j) = p(v) \hbox{  if  }
  c_{ij } (v) \neq 0 \, .
\end{equation}
For the proof of (b) we choose a basis $\{ u^i \}_{i \in S'}$ of
$\goth g_{1/2}$ such that $\langle u_i , u^j \rangle
=\delta_{ij}$ (recall that the skew-supersymmetric bilinear form
$\langle .,. \rangle$ on $\goth g_{1/2}$ defined by (2.2) is
nondegenerate).  Then we have:
\begin{equation}
  \label{eq:2.9}
  c_{ij} (v) =\langle [v,u_j], u^i \rangle \, .
\end{equation}
Furthermore, by the Jacobi identity, we have for $a,b \in \goth
g_{1/2}$ and $v \in \goth g^f_0$:
\begin{equation}
  \label{eq:2.10}
  \langle [v,a],b \rangle =(-1)^{p(a)p(b)}
  \langle [v,b],a \rangle \, .
\end{equation}
The proof of (b) is straightforward, using the Wick formula and
(\ref{eq:2.8}), (\ref{eq:2.9}), (\ref{eq:2.10}).

$\Box$ \par \vspace{.2in}

Let ${\goth h}^f$ be a maximal ad-diagonizable subalgebra of
${\goth g}^f_0$ and let ${\goth h}$ be a Cartan subalgebra of
$\goth g_0$ containing ${\goth h}^f $ (it contains $x$). We can choose a basis $\{ e_\alpha \}_{\alpha
  \in S^\prime}$ of $\goth g_{\frac{1}{2}}$ consisting of root
vectors, and extend it to a basis $\{ e_\alpha \}_{\alpha \in S}$
of ${\goth g}_+$ consisting of root vectors. Thus we may think of
$S^\prime$ and $S$ as subsets of the set of roots $\Delta \subset
\goth h^*$ of $\goth g$.

Lemma 2.1(a) implies that $[{h^{\rm ch}}_{\lambda}
\varphi_{\alpha}] = \alpha (h) \varphi_{\alpha}$, and
$[{h^{\rm ch}}_{\lambda} \varphi^*_{\alpha}] =-\alpha (h)
\varphi^*_{\alpha}$ for $h \in \goth h$ and $\alpha \in S$,
hence
\begin{equation}
  \label{eq:2.11}
  [{J^{\{ h \}}}_{\lambda} \varphi_{\alpha}]
    = \alpha (h) \varphi_{\alpha} \, , \,
       [{J^{\{ h \}}}_{\lambda} \varphi^*_{\alpha}]
    = -\alpha (h) \varphi^*_{\alpha} \hbox{  if   }
      h \in \goth h \, , \, \alpha \in S \, .
\end{equation}
Likewise, Lemma~2.1(b) implies that $[{h^{\rm ne}}_{\lambda}
\Phi_{\alpha}] = \alpha (h) \Phi_{\alpha}$ if $h \in \goth h^f$,
hence
\begin{equation}
  \label{eq:2.12}
  [{J^{\{ h \}}}_{\lambda} \Phi_{\alpha}] =\alpha (h) \Phi_{\alpha}
     \hbox{  if  } h \in \goth h^f \, , \,
     \alpha \in S' \, .
\end{equation}

Part (a) of the following theorem confirms Conjecture 2.1
in the case $j =0$.
\begin{theorem} \label{th:Jv}
\alphaparenlist
\begin{enumerate}
\item 
 If $v \in \goth g^f_0$,
  then $  [
d_\lambda J^{\{v\}} ] =0$, hence the image of each
$J^{\{v\}} \ (v \in
\goth g^f_0 ) $ is a field of the
vertex algebra $W_k (\goth g,x, f)$.

\item 
$\ [L_\lambda J^{(v)} ] = (\partial +(1-j) \lambda
) J^{(v)} + \delta_{j0}\lambda^2 (\tfrac{1}{2}\, {\rm
str}_{\goth g_+} ({\rm ad}\, v) -(k+h^{\vee})(v|x))$ if $v \in \goth
g_j$, \\
and the same formula holds for $J^{\{ v \}}$ if $v \in \goth g_0$.

\item 
$[{J^{\{v\}}}_\lambda J^{\{v^\prime \}} ] = J^{\{[v,
v^\prime ]\}} + \lambda
( k (v |
v^\prime) + {\rm str}_{{\goth g}_+}
({\rm ad} v) ({\rm ad} v^\prime) - \frac{1}{2}
{\rm str}_{{\goth g}_{\frac{1}{2}}}
({\rm ad} v) ({\rm ad} v^\prime))
\hbox  {if  }  v,v' \in {\goth g}^f_0$,
\begin{eqnarray*}
\lefteqn{\hspace{-2.85in}[{J^{(v)}}_{\lambda} J^{(v')}]
       = J^{([v,v'])}+ \delta_{i0}\delta_{j0}
    \lambda (k(v|v') + {\rm str}_{\goth g_+} ({\rm ad} v)({\rm  ad} v'))
    \hbox{ if } v \in \goth g_i,v' \in \goth g_j \hbox{ and } ij
    \geq 0 \, .}
\end{eqnarray*}
\end{enumerate}
\end{theorem}

{\it Proof.} Let $v \in \goth g_0$.  Due to (\ref{eq:2.8}) and $(\goth g_0 | \goth g_+)=0$, we obtain
from (2.4):
\begin{equation}
  \label{eq:2.13}
  [d_{\lambda} v] =-\sum_{i,j \in S} (-1)^{p(u_i)}
      c_{ij} (v) u_i \varphi^*_j \, .
\end{equation}
Next, assuming that $c_{ij} (v) \neq 0$, we compute
$[d_{\lambda}: \varphi_i \varphi^*_j:]$.  Our assumption implies
that the elements $u_i$ and $u_j$ have the same degree in the
gradation (2.3), hence the degree of their commutator is larger.
This implies that the integral term in the non-commutative Wick
formula vanishes, i.e.,~$[d_{\lambda} : \varphi_i \varphi^*_j:] =
: [d_{\lambda} \varphi_i] \varphi^*_j : + (-1)^{p(\varphi_i)} :
\varphi_i [d_{\lambda} \varphi^*_j]:$.  Therefore, by (2.4) we
obtain:
\begin{displaymath}
  [d_{\lambda}v^{\rm ch}] = \sum^5_{r=1}
  [d_{\lambda} v^{\rm ch}]_r \, ,
\end{displaymath}
where
\begin{eqnarray*}
  [d_{\lambda} v^{\rm ch}]_1 &=& -\sum_{i,j\in S}
  (-1)^{p(\varphi_i)} c_{ij}(v) : u_i \varphi^*_j:  \, , \\[1ex]
  [d_{\lambda} v^{\rm ch}]_2 &=& \sum_{i,j \in S} c_{ij}(v)
  (f|u_i) \varphi^*_j =\sum_{j \in S} (f|[v,u_j])\varphi^*_j \, ,\\[1ex]
  [d_{\lambda} v^{\rm ch}]_3 &=& \sum_{i,j,\ell,k \in S}
  (-1)^{p(u_k)} c_{ij} (v) c^{\ell}_{ik} : \varphi_{\ell}
  \varphi^*_k \varphi^*_j :\, , \\[1ex]
  [d_{\lambda} v^{\rm ch}]_4 &=& \frac{1}{2}
  \sum_{i,j,k,\ell \in  S}  (-1)^{p(u_k)p(u_j)} c_{ij}
  (v) c^j_{k\ell} : \varphi_i\varphi^*_k \varphi^*_{\ell} :\, , \\[1ex]
  [d_{\lambda} v^{\rm ch}]_5 &=&  \sum_{i,j \in S'} c_{ij}
  (v) : \Phi_i \varphi^*_j : \, .
\end{eqnarray*}
It follows from (\ref{eq:2.13}) that
\begin{equation}
  \label{eq:2.14}
  [d_{\lambda}v] + [d_{\lambda}v^{\rm ch}]_1 =0 \, ,
\end{equation}
and that
\begin{equation}
  \label{eq:2.15}
  [d_{\lambda}v^{\rm ch}]_2 = \sum_{j \in S}
  ([f,v]| u_j) \varphi^*_j \,\,(=0 \hbox{ if } v \in \goth g^f_0)\, .
\end{equation}
Furthermore, by relabeling the indices, one can write:
\begin{eqnarray*}
  [ d_{\lambda}v^{\rm ch} ]_3 &=& -\sum_{i,j,k,\ell \in S}
  (-1)^{p(u_\ell)+p(u_{\ell})p(u_k)} c_{j\ell}(v) c^i_{jk}
  : \varphi_i \varphi^*_{\ell}\varphi^*_k: \, \\
\noalign{\hbox{hence}}\\
\left[ d_{\lambda}v^{\rm ch} \right]_3
&=& \frac{1}{2} \sum_{i,j,k,\ell \in S}
    (-1)^{p(u_k)} \left(c_{j\ell}(v) c^i_{jk} -
    (-1)^{p(u_k)p(u_{\ell})} c_{jk}(v) c^i_{j\ell}\right) :
    \varphi_i \varphi^*_k \varphi^*_{\ell}: \, .
\end{eqnarray*}
Therefore
\begin{displaymath}
  [d_{\lambda}v^{\rm ch}]_3 + [d_{\lambda}v^{\rm ch}]_4 =
  \frac{1}{2} \sum_{i,\ell ,k \in S}
  (-1)^{p(u_k) +p(u_k)p(u_{\ell})} A (i,\ell ,k):
  \varphi_i \varphi^*_k \varphi^*_{\ell}:\, ,
\end{displaymath}
where $A(i,\ell ,k):= \sum_{j \in S}
((-1)^{p(u_k)p(u_{\ell})}c_{j\ell} (v) c^i_{jk} -c_{jk}(v)
c^i_{j\ell} + c_{ij} (v)
c^j_{k\ell})$.  From the Jacobi identity:
$
  [v,[u_k,u_{\ell}]] = [[v,u_k],u_{\ell}] +
  (-1)^{p(v)p(u_k)} [u_k,[v,u_{\ell}]] \, ,$
one has
\begin{displaymath}
  0= \sum_{i,j \in S} (c_{ij} (v) c^j_{k\ell}-
  c_{jk} (v) c^i_{j\ell} + (-1)^{p(u_k)p(u_{\ell})}
  c_{j \ell}(v)
  c^i_{jk}) u_i = \sum_{i \in S} A(i,\ell ,k) u_i \, ,
\end{displaymath}
which implies $A(i,\ell ,k) =0$ for all $i,\ell ,k$.  Thus we
obtain
\begin{equation}
  \label{eq:2.16}
  \left[ d_{\lambda}v^{\rm ch}\right]_3
  + \left[d_{\lambda} v^{\rm ch}\right]_4 =0 \, .
\end{equation}

Next, we compute $[d_{\lambda}v^{\rm ne}]$ for $v \in \goth
g^f_0$.  For that recall the skew-supersymmetric bilinear form
$\langle . , . \rangle$ on $\goth g_{1/2}$, given by (2.2), and
formulas (2.9) and (\ref{eq:2.10}).

Using (2.4) and (\ref{eq:2.9}), we obtain:
\begin{eqnarray}
\nonumber
       \left[ d_{\lambda}v^{\rm ne}\right]
   &=& \left[ d_{\lambda} v^{\rm ne}\right]_1 +
       \left[ d_{\lambda}v^{\rm ne}\right]_2 \, , \\
\noalign{\nonumber
\hbox{where }}\\
\nonumber
       \left[ d_{\lambda} v^{{\rm ne}}\right]_1
   &=& \frac{1}{2} \sum_{i,j,k \in S'}
       c_{ij} (v) \langle u_i,u_k \rangle \varphi^*_k \Phi_j \, , \\
\label{eq:2.17}
       \left[ d_{\lambda}v^{{\rm ne}}\right]_2
   &=& - \frac{1}{2} \sum_{i,j \in S'} c_{ij} (v)
            \Phi_i \varphi^*_j \, .
\end{eqnarray}
We have:
\begin{displaymath}
  [d_{\lambda}v^{\rm ne}]_1 = \frac{1}{2} \sum_{i,j,k}
  (-1)^{p(u_j)(p(u_k)+1)} c_{ij}(v)
  \langle u_i,u_k \rangle \Phi^j \varphi^*_k \, .
\end{displaymath}
Using that $\Phi^j =\sum_{r \in S'} \langle u^j,u^r \rangle
\Phi_r$ and that $p(u_i)=p(u_k)$ if $\langle u_i,u_k \rangle \neq
0$, we obtain:
\begin{displaymath}
  [d_{\lambda}v^{\rm ne}]_1 = -\frac{1}{2} \sum_{i,j,k,r}
  (-1)^{p(u_i)p(u_j)} c_{ij}(v) \langle u_i,u_k \rangle
  \langle u^r,u^j \rangle \Phi_r \varphi^*_k \, .
\end{displaymath}
Using (2.9) and (\ref{eq:2.10}), we obtain:
$(-1)^{p(u_i) p(u_j)} c_{ij}(v) = \langle [v,u^i],u_j \rangle$,
hence:
\begin{displaymath}
  [d_{\lambda} v^{\rm ne}]_1 = -\frac{1}{2} \sum_{k,r}
  \langle [v, \sum_i \langle u_i,u_k \rangle u^i] \, , \,
  \sum_j \langle u^r,u^j \rangle u_j \rangle \Phi_r \varphi^*_k
  =-\frac{1}{2} \sum_{k,r} c_{rk} (v) \Phi_r \varphi^*_k \, ,
\end{displaymath}
and, by (\ref{eq:2.17}), we
obtain:
\begin{displaymath}
  [d_{\lambda} v^{\rm ne}] =-\sum_{i,j \in S'}
  c_{ij}(v) \Phi_i \varphi^*_j \, .
\end{displaymath}
Thus, we see that for $v \in \goth g^f_0$ one has
\begin{equation}
  \label{eq:2.18}
  [d_{\lambda} v^{\rm ch}]_5 + [d_{\lambda} v^{\rm ne}]=0 \, .
\end{equation}
Comparing (\ref{eq:2.14}), (\ref{eq:2.15}), (2.16) and
(\ref{eq:2.18}) gives (a).

By (1.10), one has
\begin{displaymath}
  [L_{\lambda}v] = (\partial + \lambda)v-\lambda^2 k(x|v) \, ,
\end{displaymath}
and by Theorem~2.2(b) and the noncommutative Wick
formula the following relations hold:
\begin{displaymath}
  [L_{\lambda}: \varphi_i\varphi^*_j:] = (\partial + \lambda):
  \varphi_i \varphi^*_j :+ \left( \tfrac{1}{2} -m_i \right)
  \delta_{i,j} \lambda^2 \, , \quad
  [L_{\lambda}: \Phi_i \Phi^j:] = (\partial +\lambda):
  \Phi_i \Phi^j : \, .
\end{displaymath}
Hence:
\begin{displaymath}
  [L_{\lambda} v^{\rm ch}] = (\partial + \lambda) v^{\rm ch}
     + \lambda^2 \Large( \tfrac{1}{2} {\rm str}_{\goth g_+}
       {(\rm ad} \, v) -\sum_{i \in S} (-1)^{p(u_i)}
       m_ic_{ii} (v) \Large) \, , \quad
     [L_{\lambda} v^{\rm ne}] = (\partial + \lambda) v^{\rm ne}
     \, .
\end{displaymath}
As $\sum_{i \in S} (-1)^{p(u_i)}m_ic_{ii}(v) =h^{\vee} (x|v)$ by
(1.7), (b) follows.

The proof of (c) is similar.  It uses only the usual Wick
formula.  We omit the details.

$\Box$ \par \vspace{.2in}


\subsection{ Construction of the $ W_k
( \goth g,x, f )$-modules }
 Let $M$ be a restricted
$\hat{\goth g}$-module of level $k$
(i.e. $K = k \ I_M$). It extends to the $V_k (
\goth g )$-module, and then to the
${\cal C} ( \goth g,x,  f, k )$-module
$$
{\cal C} ( M ) = M \bigotimes F ( \goth g,x,  f ) \ .
$$
One has the charge decomposition of ${\cal C} ( M )$ induced
by that of $F( \goth g, x, f )$ by setting the charge of $M$ to
be zero:
$$
{\cal C} ( M ) = \bigoplus_{m \in \ZZ} {\cal C} ( M )_m \ .
$$
Furthermore, $({\cal C} ( M ), d_0 )$ form a  ${\cal
  C}( \goth g,x, f , k )$-module complex, hence its homology, $H ( M
)=\oplus_{j\in \ZZ}$ $ H_j( M )$, is a direct sum of $W_k ( \goth g,
x, f )$-modules. We thus get a functor, which we denote by $H$,
from the category of restricted $\hat{\goth g}$-modules to the
category of $\ZZ$-graded $W_k (\goth g, x, f )$-modules, that send $M$ to $H (M )$.

{\bf Remark 2.3.}  Let $\vac$ be the vacuum vector of the vertex
algebra $F(\goth g ,x,f)$ and let $v \in M$ be such that $(\goth
g_+ t^m)(v) =0$ for all $m \geq 0$.  Then
\begin{displaymath}
  d_0 (v \otimes \vac ) =0 \, .
\end{displaymath}
In particular if $M$ is a highest weight
$\hat{\goth g}$-module with highest weight $\Lambda$ of level $k
\neq -h^{\vee}$, and $v_{\Lambda}$ is the highest weight vector,
then $d_0 (v_{\Lambda} \otimes \vac )=0$.  So, if the vector
$v_{\Lambda} \otimes \vac$ is not in the image of $d_0$, its
image in $H_0 (M)$, which we denote by $\tilde{v}_{\Lambda}$,
generates a non-zero $W_k (\goth g ,x,f)$-submodule.  Its central charge
is given by formula~(2.6).  The eigenvalue of $L_0$ on
$\tilde{v}_{\Lambda}$ is equal to (cf.~Section~3.1):
\begin{equation}
\label{eq:2.19}
  \frac{(\Lambda | \Lambda +2\hat{\rho})}{2(k+h^{\vee})}
   \, - \, (x+D|\Lambda) \, .
\end{equation}
The eigenvalue of $J^{\{ h \}}_0$ $(h \in \goth h^f)$ on
$\tilde{v}_{\Lambda}$ is equal to $\Lambda (h)$.

\section{Character Formulas}
\setcounter{equation}{0}
\subsection{The Euler-Poincar\'{e} character of
$H(M)$  }
Let $\goth g$ be one of the
basic simple finite-dimensional Lie superalgebras.
Recall that, apart from the five exceptional Lie algebras, they are as follows:
$s\ell (m|n) / \delta_{m,n}
\CZ I$, $osp (m | n)$, $D(2,1; a)$,
$F(4)$ and $G(3)$ \cite{K1}. Recall that
$\goth g$ carries a unique (up to a
constant factor) non-degenerate invariant
bilinear form
\cite{K1}, and it is automatically even
supersymmetric. We choose one of them, and
denote it by $( . | . )$.

Given $h \in {\goth h}^f$, define, as
before, the fields $h^{\rm ch}(z)$ and $h^{\rm
ne}(z)$. They are given by the following slightly simpler formulas:
\begin{eqnarray*}
h^{\rm ch} (z) = - \sum_{\alpha \in S}
\alpha (h) : \varphi_\alpha^\ast (z)
\varphi_\alpha (z) : \ , &
h^{\rm ne} (z) & = - \frac{1}{2}
 \sum_{\alpha
\in S^\prime} \alpha (h) : \Phi^\alpha (z)
\Phi_\alpha (z) : \ .
\end{eqnarray*}
Since these fields are of conformal weight 1, we
write: $h^{\rm ch} (z) = \sum_{n \in \ZZ} h^{\rm
ch}_n z^{-n-1} $, and $ h^{\rm ne} (z) = \sum_{n \in
\ZZ} h^{\rm ne}_n z^{-n-1}$.  Likewise, we write $J^{\{ h \}} (z)
= \sum_{n \in \ZZ} J^{\{ h \}}_n z^{-n-1}$.

Let $\hat{\goth h} =
{\goth h} + \CZ K + \CZ D$ be the Cartan
subalgebra of the affine Lie superalgebra $\hat{\goth
g}$. As usual, we extend a  root $\alpha  \in
{\triangle}$ to
${\goth h}$ by letting $\alpha ( K ) = \alpha ( D
) = 0$. We extend the bilinear form $(. | . )$
from ${\goth h}$ (on which it is
non-degenerate) to $\hat{\goth h}$ by letting:
$$
( {\goth h} | \CZ K + \CZ
D ) = 0 \ , \ \ \ (K | K ) = ( D | D ) = 0 \ , \
\ \ ( K | D ) = 1 \ .
$$
We shall identify $\hat{\goth h}$ with $\hat{\goth h}^*$
via this form. The bilinear form $( . | . )$
extends further to the whole $\hat{\goth g}$ by
letting $( t^m a | t^n b ) = \delta_{m, -n} ( a |
b )$. Let $\hat{\Omega}$ be the Casimir operator
for $\hat{\goth g}$ and this bilinear form. Recall
that its eigenvalue for a  $\hat{\goth g}$-module
with the highest weight $\Lambda$ is equal to
$(\Lambda | \Lambda + 2 \hat{\rho} )$ \cite{K3}.
Denote by
$\hat{\triangle} \subset  \hat{\goth h}^* = \hat{\goth h}$ the
set of roots of $\hat{\goth g}$ with respect to
$\hat{\goth h}$.

Recall that $\hat{\triangle} =
\hat{\triangle}^{\rm re} \cup \hat{\triangle}^{\rm im}$, where
$$
\hat{\triangle}^{\rm re} = \{ \alpha + n K \ | \ \alpha
\in {\triangle} , \ n \in \ZZ \
\} \ , \ \ \ \hat{\triangle}^{\rm im} = \{ n K
\ |  \ n \in \ZZ \setminus \{ 0 \} \
\}
$$
are the sets of real and imaginary roots
respectively. Choosing a set of positive roots
${\triangle}_{0 +}$ of the set of roots
${\triangle}_0 = \{ \alpha
\in{\triangle} \ | \ ( \alpha | x) = 0
\}$, we get a set of positive roots
${\triangle}_+ = \{ \alpha
\in{\triangle} \ | \ ( \alpha | x) > 0 \}
\cup {\triangle}_{0 +} $ of
$\goth g$ and the set of positive roots
$$
\hat{\triangle}_+ = {\triangle}_+  \bigcup \{
\alpha + n K \ | \ \alpha \in
{\triangle} \cup \{ 0\} , \ n > 0
\}
$$
of $\hat{\goth g}$. We shall denote by
$\hat{\triangle}_{\rm even}$ and $ \hat{\triangle}_{\rm
odd}$, $\hat{\triangle}_{+ \rm even}$ and $
\hat{\triangle}_{+ \rm odd }$,
 etc. the sets of even and odd roots respectively.

Introduce the following subsets of
$\hat{\triangle}$ (where $n \in \ZZ$):
$$
\begin{array}{ll}
\widehat{S} = & \{ \alpha + n K \ | \ \alpha
\in S \ , \ n \geq 0 \} \bigcup \{  - \alpha + n
K \ | \ \alpha \in S \ , \ n > 0 \} \ ,
\\[1ex]
\widehat{S}^\prime =& \{ -\alpha + n K \ | \
\alpha
\in S^\prime \ , \ n > 0 \} \ .
\end{array}
$$

As usual, we write $L(z) =
\sum_{n \in \ZZ} L_n z^{-n-2}$, $L^{\goth g}(z) =
\sum_{n \in \ZZ} L_n^{\goth g} z^{-n-2}$, etc
(see Section 2.2).  Recall that we have
\cite{K3}:
$$
L^{\goth g}_0 = \frac{\hat{\Omega}}{2 (k + h^\vee)} - D
$$
for any highest weight $\hat{\goth g}$-module $M$ of
level $k$,  $ k \neq - h^\vee$.

We shall
coordinatize
$\hat{\goth h}$ by letting
$$
( \tau , z , u ) = 2 \pi i
( z -
\tau D + u K ) \ ,
$$
where $z \in {\goth h}$, $\tau, u \in
\CZ$. We shall assume that ${\rm
Im} \ \tau > 0$ in order to guarantee the
convergence of  characters, and set $q = e^{2
\pi i \tau}$. Define the character of a $\hat{\goth g}$-module
$M$ by ${\rm ch}_M := {\rm tr}_M \ e^{2 \pi
i ( z -
\tau D + u K )}$. For any highest
weight
$\hat{{\goth g}}$-module $M$ of level $k \neq - h^\vee$
the series  ${\rm ch}_M$ converges to an analytic
function in the interior of the domain $ Y_> :=\{ h \in
\hat{\goth h} \ | \  ( \alpha | h ) > 0 \ {\rm for \
all } \  \ \alpha \in \hat{\triangle}_+ \} $; moreover,
the domain of convergence $Y(M)$ is a convex
domain contained in the upper half space $Y = \{  h \in
\hat{\goth h} \ | \ {\rm Re} ( h | K ) > 0 \} =
\{ ( \tau , z , u ) \ | \ {\rm Im} \tau > 0 \}$
(\cite{K3}, Lemma 10.6).

\begin{lemma} \label{lem:trF}
For a regular element $b \in {\goth h}$
(i.e.,~$ \alpha ( b ) \neq 0$ for all $\alpha \in
{\triangle}$), $h \in {\goth h}^f$  and  any sufficiently small
$\epsilon \in \CZ \setminus \{0\}$, one has the
following formula for the Euler-Poincar\'{e}
character of
$F (\goth g, x,f)$:
\be
\sum_{j \in \ZZ} (-1)^j {\rm tr}_{F_j} \bigg(
q^{L_0^{\rm ch} + \epsilon b_0^{\rm ch}+
L_0^{\rm ne} } e^{2 \pi i (h_0^{\rm ch} +h_0^{\rm ne})}
\bigg)=
\prod_{\alpha
\in
\widehat{S}
\setminus
\widehat{S}^\prime} ( 1- s (\alpha)
e^{-\alpha})^{s(\alpha)} ( \tau, \tau (\epsilon b
- x) + h , 0) \ ,
\ele(chM)
where $s ( \alpha ) := (-1)^{p(\alpha)} , \alpha
\in \hat{\triangle}$.
\end{lemma}

{\it Proof.}  Since the fields $\varphi^*_{\alpha}$ and
$\varphi_{\alpha}$ (resp.~$\Phi_{\alpha}$) are primary with
respect to $L^{\rm ch}$ (resp.~$L^{\rm ne}$) of conformal weights
$(\alpha |x)$ and $1-(\alpha |x)$ (resp.~$\frac{1}{2}$), we have:
\begin{eqnarray*}
  [L^{\rm ch}_0 \, , \, \varphi_{\alpha (-n)}] &=&
     (n-(\alpha | x)) \varphi_{\alpha (-n)} \, ,\\
{}  [L^{\rm ch}_0 \, , \, \varphi^*_{\alpha (-n)}] &=&
     (n-1+(\alpha |x)) \varphi^*_{\alpha (-n)} \, ,\\
 {} [L^{\rm ne}_0 \, , \, \Phi_{\alpha (-n)}] &=&
     (n-\frac{1}{2}) \Phi_{\alpha (-n)} \, .
\end{eqnarray*}
Using this, we get:
\begin{eqnarray*}
  \sum_{j \in \ZZ} (-1)^j {\rm tr}_{F(A)_j}q^{L^{\rm ch}_0}
     &=& \prod_{\alpha \in S} \prod^{\infty}_{n=1}
        \left( 1-s(\alpha)q^{(nK-\alpha|D+x)}\right)^{s(\alpha)} \, ,\\
  \sum_{j \in \ZZ} (-1)^j {\rm tr}_{F(A^*)_j}q^{L^{\rm ch}_0}
     &=& \prod_{\alpha \in S} \prod^{\infty}_{n=1}
        \left( 1-s(\alpha)q^{((n-1)K+\alpha |D+x)}
        \right)^{s(\alpha)}\, ,\\
   {\rm tr}_{F(A_{\rm ne})}q^{L^{\rm ne}_0}
     &=& \prod_{\alpha \in S'} \prod^{\infty}_{n=1}
       \left( 1-s(\alpha)q^{(nK-\alpha
           |D+x)}\right)^{-s(\alpha)}\, .
\end{eqnarray*}

Using these formulas along with (2.11) and
(2.12), we get for $h \in \goth h$:
\begin{eqnarray*}
 \sum_{j \in \ZZ}(-1)^j {\rm tr}_{F(A)_j} q^{L^{\rm ch}_0}
    e^{2\pi i J^{\{ h \}}_0} &=&
    \prod_{\alpha \in S} \prod^{\infty}_{n=1}
    \left(1-s(\alpha)e^{2\pi i (-nK+\alpha|-\tau (D+x)+h)}
      \right)^{s(\alpha)} \, \\[1ex]
  \sum_{j \in \ZZ} (-1)^j {\rm tr}_{F(A^*)_j}
    q^{L^{\rm ch}_0}e^{2\pi i J^{\{ h \}}_0} &=&
    \prod_{\alpha \in S} \prod^{\infty}_{n=1}
    \left( 1-s(\alpha)e^{2\pi i (-(n-1)K-\alpha|-\tau (D+x)+h)}
      \right)^{s(\alpha)} \, , \\[1ex]
\end{eqnarray*}
and for $h \in \goth h^f$:
\begin{displaymath}
  {\rm tr}_{F(A_{\rm ne})} q^{L^{\rm ne}_0} e^{2\pi i J^{\{ h \}}_0}
   = \prod_{\alpha \in S'} \prod^{\infty}_{n=1}
   \left(1-s(\alpha ) e^{2\pi i (-nK+\alpha| -\tau
       (D+x) +h)}
     \right)^{-s(\alpha)} \, .
\end{displaymath}
The lemma follows immediately from the last three identities.

$\Box$ \par \vspace{.2in}

Note that the right hand side of (3.1)
defines a meromorphic function on $Y$ with
simple poles on the hyperplanes $T_{\alpha} : = \{ h \in
  \hat{\goth h} |\,  \alpha (h) =0 \} \, , \,
\alpha \in \hat{\triangle}^{\rm re}_{\rm even}$.

Let $M$ be a highest weight $\hat{\goth g}$-module
of level $k \neq - h^\vee$. We shall
assume that its character ${\rm ch}_M$ extends to
a meromorphic function in the whole upper half
space $Y$ with at most simple poles at the
hyperplanes $T_{\alpha}$, where $
\alpha \in \hat{\triangle}^{\rm re}_{\rm even}$. (We
conjecture that this is always the case.)

Let $H (M)$ be the $W_k (\goth g,x,
f)$-module defined in Section 2.5. Define the
Euler-Poincar\'{e} character of $H (M)$:
$$
{\rm ch}_{ H (M) }(h) = \sum_{j \in \ZZ} (-1)^j {\rm
tr}_{H_j(M)} q^{L_0} e^{2 \pi i J_0^{\{h\}}} ,
$$
where $h \in {\goth h}^f$ (see
Theorem~2.4). We have the following
formula for this character:
\be
{\rm ch}_{H (M)} (h)  = q^{\frac{\hat{\Omega}  |  M  }{2(k+h^\vee)}}
\lim_{\epsilon \rightarrow 0} \bigg( {\rm ch}_M
\prod_{\alpha \in \widehat{S} \setminus
\widehat{S}^\prime } ( 1 - s (\alpha ) e^{-
\alpha} )^{s (\alpha) } \bigg) ( \tau, \tau (
\epsilon b - x ) + h , 0 ) \ .
\ele(EHM)
Indeed, by the Euler-Poincar\'{e} principle we
have
$$
\begin{array}{ll}
{\rm ch}_{H (M) }(h) &= \lim_{ \epsilon \rightarrow 0}
\sum_{j
\in
\ZZ} (-1)^j {\rm tr}_{{\cal C}_j(M)} q^{L_0 +
\epsilon ( b + b_0^{\rm ch} ) }e^{2 \pi i
J_0^{\{h\}}} \\[1ex]
 &= q^{\frac{ \hat{\Omega} |
M }{2(k+h^\vee)}}
\lim_{ \epsilon
\rightarrow 0} \{ {\rm tr}_M q^{-D + \epsilon b -x}
e^{2 \pi  i h}
\sum_{j \in \ZZ} (-1)^j {\rm tr}_{F_j}
q^{L_0^{\rm ch} +
\epsilon  b_0^{\rm ch} + L_0^{\rm ne} } e^{2 \pi
i (h_0^{\rm ne} +h^{\rm ch}_0)}\} \ .
\end{array}
$$
Now (3.2) follows from Lemma~3.1.

Introduce the Weyl denominator
$$
\hat{R} = \prod_{\alpha \in \hat{\triangle}_+} ( 1- s(\alpha)
e^{-\alpha})^{s(\alpha){\rm mult} \alpha} \ .
$$
Rewriting the RHS of (3.2) using $\hat{R}$, we arrive at the
following result.

\begin{theorem} \label{th:3.1}
Let $M$ be the highest weight ${\hat{\goth g}}$-module
with the highest weight $\Lambda$ of level $k
\neq - h^\vee$, and suppose that ${\rm ch}_M$
extends to a meromorphic function on $Y$ with at
most simple poles at the hyperplanes $T_\alpha$, where
$\alpha \in \hat{\triangle}^{\rm re}_{\rm even}$. Then
\bea(ll)
 {\rm ch}_{H(M)}(h) = & \frac{
q^{ \frac{(\Lambda | \Lambda + 2\hat{
\rho})}{2(k+h^\vee)} } }
{ \prod_{j=1}^\infty
(1-q^j)^{{\rm dim} {\goth h}}} \ (\hat{R} \,
{\rm ch}_M ) (H) \\
 & \times
\prod_{n=1}^\infty
\prod_{ \alpha \in
\triangle_+, ( \alpha | x )=  0 } \bigg( ( 1 - s(\alpha) e^{-(n-1)K
-\alpha} )^{- s(\alpha)} ( 1 - s(\alpha) e^{-n K
+\alpha} )^{- s(\alpha)} \bigg)  (H) \, , \\
 & \times \prod^{\infty}_{n=1} \prod_{\alpha \in \Delta_+ ,
   (\alpha |x)=\frac{1}{2}}
 (1-s(\alpha)e^{-nK+\alpha})^{-s(\alpha)}(H) \, ,
\elea(chM)
where, as before, $s(\alpha) = (-1)^{p(\alpha)}$
and
$H := (\tau, -\tau x + h, 0 ) =  2
\pi  i ( - \tau D  - \tau x + h  ) $, $h \in \goth h^f$.
\end{theorem}
{\bf Remark 3.1.} Here is a slightly more explicit
expression for ${\rm ch}_{H(M)}$:
\begin{eqnarray*}
& {\rm ch}_{H(M)}(h) =  \frac{
q^{ \frac{(\Lambda | \Lambda + 2
\hat{\rho})}{2(k+h^\vee)} } }{ \prod_{j=1}^\infty
(1-q^j)^{{\rm dim} {\goth h}}} \ ( \hat{R}\,
{\rm ch}_M ) (\tau , - \tau x + h , 0)
\\
&\times \prod_{n=1}^\infty  \prod_{ \alpha \in
{\triangle}_+ , ( \alpha | x )=
\frac{1}{2} }
 (1 - s(\alpha) q^{n-\frac{1}{2} } e^{2
\pi i(\alpha | h )} )^{-s(\alpha)}   \\
& \times  \prod_{n=1}^\infty \prod_{ \alpha \in
{\triangle}_+, ( \alpha | x )=
0 } ( 1 - s(\alpha) q^{n-1} e^{-2 \pi
i(\alpha | h )} )^{-s(\alpha)} ( 1 - s(\alpha)
q^n e^{2
\pi i(\alpha | h )} )^{-s(\alpha)} \, .
\end{eqnarray*}
Since we may assume that $(\gamma_i |x) \geq 0$, for a set of simple roots $\{ \gamma_i \}$ of $\Delta_{+ {\rm even}}$, is easy to show that if the set $\{ \alpha \in \Delta_{+{\rm even}}| (\alpha |x)=0 \}$ is non-empty, then the restriction of each $\alpha$ from this set to $\goth h^{f}$ is a non-zero linear function.

 \subsection{ Conditions of non-vanishing of $
H(M) $ } Using Theorem~3.1, we can
establish a necessary and sufficient condition for  ${\rm ch}_{H(M)}$
to be not
identically zero, hence a sufficient condition for the
non-vanishing of $H(M)$.

\begin{theorem} \label{th:3.2}
Let $M$ be as in Theorem~3.1. Then
${\rm ch}_{H(M)}$ is not identically zero if and only
if the $\hat{\goth g}$-module $M$ is not locally
nilpotent with respect to all root spaces ${\goth
g}_{-\alpha}$, where $\alpha$ are positive even
real roots satisfying the following three
properties:
$$
(i) \ ( \alpha | D + x ) = 0 , \ \ \ (ii) \
 (\alpha | {\goth h}^f) = 0 , \ \ \
(iii) \ | ( \alpha |  x ) | \geq 1 \ .
$$
In particular, these conditions guarantee that $H (M) \neq
0$.
\end{theorem}

\begin{lemma} \label{lem:}
Let $\alpha \in \hat{\triangle}^{\rm re}_{+ {\rm
even}}$. Then the function ${\rm ch}_M$ is
analytic on a non-empty open subset of the
hyperplane $T_\alpha$ if and only if $\hat{{\goth
g}}_{-\alpha}$ is locally nilpotent on $M$.
\end{lemma}
{\it Proof.} If $\hat{\goth
g}_{-\alpha}$ is locally nilpotent on $M$, then
$r_\alpha {\rm ch}_M = {\rm ch}_M$, where
$r_\alpha$ is a reflection with respect to the
hyperplane $T_\alpha$
\cite{K3}. Hence $Y(M)$ is an
$r_\alpha$-invariant convex domain and therefore,
$Y (M) \cap T_\alpha$ contains a non-empty open
set (it is because any segment connecting $a$ and $ r_\alpha a
$, where $a \in Y_>$, has a non-empty intersection
with the hyperplane $T_\alpha$ ).

Conversely, suppose that $\hat{\goth g}_{-\alpha}$ is
not locally nilpotent on $M$. Consider $
s\ell_2 \subset \hat{\goth g}$ generated by $\hat{\goth
g}_{-\alpha}$ and $\hat{\goth g}_{\alpha}$, and let
$M_{\rm int}$ denote the subspace of $M$
consisting of locally finite vectors with
respect to this  $\goth{sl}_2$. Then ${\rm
ch}_{M_{\rm int}}$ is $r_\alpha$-invariant, hence
(as above) it is analytic on an open subset of
$T_\alpha$. On the other hand, ${\rm
ch}_{M/M_{\rm int}}$ is a sum of functions of the
form $\frac{e^\lambda}{1- e^{-\alpha}}$, where
$\lambda
$ is a weight of $M$. Hence ${\rm ch}_M = {\rm
ch}_{M_{\rm int}} + \frac{f}{1- e^{-\alpha}}$,
where $f$ is a meromorphic function on $Y$, which
is analytic and non-zero on a non-empty open subset
of $T_\alpha $.
$\Box$ \par \vspace{.2in}
{\it Proof of Theorem~3.2.} It
follows from Theorem~3.1 and Lemma~3.2 that ${\rm ch}_{H(M)}$ is not identically zero
if and only if $\hat{R} {\rm ch}_M$ cannot be
decomposed as the product of $1 - e^{-\alpha}$
and a meromorphic function which is analytic in a
non-zero open subset of $T_\alpha$ for each
positive even real root $\alpha$ such that $(
\alpha | H ) = 0 $
and $\alpha \not\in \{ nK -\gamma | (\gamma |x)=0$ or
$\frac{1}{2}\}  \cup \{ nK +\gamma | (\gamma |x)=0 \} $. But $(
\alpha | H ) = 2 \pi i ( - \tau (\alpha | D
+ x ) + ( \alpha| h))$, hence $(\alpha |H)=0$ is equivalent to
$(i)$ and $(ii)$.  The second condition on $\alpha$ is equivalent
to $(iii)$. Hence $ {\rm ch}_{H(M)}$ is not identically zero
if and only if  conditions $(i)$---$(iii)$
hold.
$\Box$ \par \vspace{.2in}

A $\hat{\goth g}$-module $M$ is called \emph{non-degenerate} if each
$\hat{\goth g}_{-\alpha}$, where $\alpha$ is a positive real even root
satisfying properties (i)---(iii) (in Theorem~3.2), is not locally nilpotent on $M$.
Otherwise $M$ is called \emph{degenerate}.

\subsection{Admissible highest weight $\hat{\goth g}$-modules}
\label{sec:3.3}

Fix a non-degenerate invariant bilinear form $(.|.)$ on $\goth g$
such that all $(\alpha | \alpha) \in \RR$ for $\alpha \in
\Delta$.  Then we have a decomposition of the set of even roots
$\Delta_{\overline{0}}$ into a disjoint union of
$\Delta^>_{\overline{0}}$ and $\Delta^<_{\overline{0}}$, where
$\Delta^>_{\overline{0}}$ (resp. $\Delta^<_{\overline{0}}$) is
the set of $\alpha \in \Delta_{\overline{0}}$ such that $(\alpha
| \alpha)>0$ (resp. $<0$).  Let $\goth g^>_{\overline{0}}$ be the
semisimple subalgebra of the reductive Lie algebra $\goth
g_{\overline{0}}$ with root system $\Delta^>_{\overline{0}}$, and
let $\hat{\goth g}^>_{\overline{0}}$ be the affine subalgebra of
$\hat{\goth g}$ associated to $\goth g^>_{\overline{0}}$.

Recall that a $\hat{\goth g}$-module $L(\Lambda)$ is called
\emph{integrable} if it is integrable with respect to $\hat{\goth
  g}^>_{\overline{0}}$ and is locally finite with respect to
$\goth g$.  In \cite{KW4} a complete classification of integrable
$\hat{\goth g}$-modules was obtained.

\textbf{Definition.} (\cite{KW1}, \cite{KW2}, \cite{KW4}).
 Let $\hat{\Delta}'
\subset \hat{\Delta}$ be a subset such that $\QQ\hat{\Delta}'=\QQ
\hat{\Delta}$ and $\hat{\Delta}'$ is isomorphic to a set of roots of an affine
superalgebra $\hat{\goth g}'$ (which is not necessarily a
subalgebra of $\hat{\goth g}$).  Let $\hat{\Pi}' \subset
\hat{\Delta}_+$ be the set of simple roots of $\hat{\Delta}'$
(for the subset of positive roots $\hat{\Delta} \cap
\hat{\Delta}_+$).  Let $\hat{\rho}' \in \goth h'$ be the Weyl
vector, i.e.,~$2 (\hat{\rho}'|\alpha')=(\alpha' |\alpha') $
for all $\alpha' \in \hat{\Pi}'$.
A $\hat{\goth g}$-module $L(\Lambda)$ (and the weight
$\Lambda$) is called \emph{admissible} for $\Delta'$ if the
$\hat{\goth g}'$-module $L' (\Lambda +\hat{\rho} -\hat{\rho}')$  is
integrable and this condition does not hold for any $\hat{\Delta''}
\supsetneqq \hat{\Delta'}$.  It is called \emph{principal admissible} if
$\hat{\Delta}'$ is isomorphic to $\hat{\Delta}$.

{\bf Conjecture 3.3A.}\cite{KW2}, \cite{KW4}   The character of an admissible
$\hat{\goth g}$-module $L (\Lambda)$ is related to the character
of an integrable $\hat{\goth g}'$-module by the formula:
\begin{equation}
  \label{eq:3.4}
e^{\hat{\rho}} \hat{R} {\rm ch}_{L(\Lambda)} =e^{\hat{\rho}'}
    \hat{R}' {\rm ch}_{L'(\Lambda + \hat{\rho}-\hat{\rho}')} \, .
\end{equation}

{\bf Remark 3.3.}  Formula (\ref{eq:3.4}) holds for general
symmetrizable Kac--Moody Lie algebra.  It is immediate from the
character formula for admissible modules \cite{KW1}, \cite{KW2}.  In fact
(\ref{eq:3.4}) holds for these Lie algebras in the much more
difficult case when ``integrable'' is replaced by ``integral'' \cite{F}.

\textbf{Definition.} \cite{KW4}.  A $\hat{\goth g}$-module $L(\Lambda)$
is called \emph{boundary admissible} for $\hat{\Delta}'$ if
$\Lambda +\hat{\rho} - \hat{\rho}' =0$ (i.e.,~dim$L'(\Lambda
+\hat{\rho}-\hat{\rho}')=1)$.

Of course, (\ref{eq:3.4}) provides an explicit product formula
for the boundary admissible $\hat{\goth g}$-modules:
\begin{equation}
  \label{eq:3.5}
  {\rm ch}_{\Lambda} =e^{\Lambda} \hat{R}'/\hat{R} \,.
\end{equation}


{\bf Conjecture 3.3B.} If $L(\Lambda)$ is an admissible $\hat{\goth
g}$-module, then the $W_k (\goth{g} ,x,f)$-module $H(L(\Lambda))$
is either zero or irreducible.

If Conjecture~3.3B holds, then
Theorem~3.2 gives necessary and sufficient conditions for the vanishing of $H
(L (\Lambda))$.


\section{Vertex Algebras $W_k (\goth g,
  e_{-\theta})$, where $\theta$ is a Highest Root }
\setcounter{equation}{0}
We now choose a subset of positive roots
in the set of roots ${\triangle}$ such that the highest root
$\theta$ (i.e.,~$\theta + \alpha$ is not a root for any positive
root $\alpha$) is even. In this section, we shall classify all
the examples of vertex algebras $W_k (\goth g , f)$ where $f = e_{-\theta}$. Denote by
$e=e_\theta$ the root vector such that $(e | f )
= (\theta | \theta)^{-1}$.
Let
$x = \frac{\theta}{(\theta |
\theta)}$, so that $\theta ( x ) = 1$. Then $\langle e,x,f
\rangle$  is an $s\ell_2$-triple.  Furthermore, we  have :
\be
S = S^\prime \cup \{ \theta \} \ .
\ele(Stheta)
Indeed, otherwise there exists an element $\alpha
\in\triangle \backslash \{ \theta \}$ such that
$\frac{ 2 \, (\alpha |
\theta )}{( \theta | \theta )} \geq 2$, hence
$\alpha - 2 \,\theta \in \triangle$.
This is impossible since $\alpha - 2 \theta < - \theta$.

Thus, the $\frac{1}{2} \ZZ$-gradation (2.1) of ${\goth g}$
has the form:
\begin{equation}
  \label{eq:4.2}
 {\goth g} =  {\goth g}_{-1} + {\goth g}_{-\frac{1}{2}}
+{\goth g}_0
 + {\goth g}_{\frac{1}{2}} + \goth g_1, \hbox { where }
 \goth g_{-1} =\CC f, \goth g_1 = \CC e \, .
\end{equation}
One also has:
\begin{equation}
  \label{eq:4.3}
  \goth g^f = \goth g_{-1} + \goth g_{-\frac{1}{2}} +
  \goth g^f_0 \, , \,
  \goth g_0 = \goth g^f_0 \oplus \CC x \, ,
\end{equation}
where
\begin{displaymath}
  \goth g^f_0 = \{ a \in\goth g_0 | (a|x)=0 \}
      = \goth h^f \oplus (\oplus_{\alpha \in \Delta_0} \CC
      e_{\alpha})\, ,\,
\goth h^f = \{  h \in \goth h | (h|x)=0 \} \, ,
    \Delta_0 = \{ \alpha \in \Delta | (\alpha |x)=0 \} \, .
\end{displaymath}

It is easy to see now that formula~(2.5) for the central
charge of the Virasoro algebra of $W_k (\goth g,e_{-\theta})$
becomes:
\begin{equation}
  \label{eq:4.4}
  c=\frac{k}{k+h^{\vee}} {\rm sdim} \, \goth g -
  \frac{12k}{(\theta | \theta)} +\frac{1}{4}
  ({\rm sdim} \,\goth g -{\rm sdim}\, \goth g^f_0)-\frac{11}{4} \, .
\end{equation}

Furthermore, it is easy to see from Theorem~2.4(b) that all
fields $J^{\{v\}}$, $v \in\goth g^f_0$, of the vertex algebra $W_k
(\goth g , e_{-\theta})$ are primary (of conformal weight $1$). Indeed, we have for $v\in
\goth g^f_0$:
\begin{displaymath}
  {\rm str}_{\goth g_+} \ad \, v =\frac{1}{2} {\rm str}_{\goth g}
  (\ad \, v) (\ad \, x) = h^{\vee} \, (v|x)
\end{displaymath}
by (1.7).
But $(v|x) =-(v|[f,e]) =- ([v,f]|e)=0$.  (Hence $(\goth g^f |x) =0$ in any Dynkin gradation.)

Likewise, by Theorem~2.4(c),  the $2$-cocycle
of the affine subalgebra $({\goth g}^f_0)^{\hat{}}$ of $W_k (\goth g ,
e_{-\theta})$ equals:
\begin{equation}
  \label{eq:4.5}
  \alpha(v,v')=  k(v|v') + \frac{1}{2}h^{\vee} (v|v') -\frac{1}{4}
      {\rm str}_{\goth g_0} (\ad \,v) (\ad \, v') \, .
\end{equation}
In the case when $\goth g^f_0$ simple, denoting by $h^{\vee}_0$
its dual Coxeter number for $(.|.)$ restricted to $\goth g^f_0$,
we can rewrite (\ref{eq:4.5}):

\begin{equation}
  \label{eq:4.6}
  \alpha (v,v')=(v|v') (k + \frac{1}{2} (h^{\vee}-h^{\vee}_0)) \, .
\end{equation}

The following proposition lists all vertex algebras $W_k (\goth g ,e_{-\theta})$.

\begin{proposition}  \label{prop:glist}
All cases of $({\goth g}, \theta )$
along with the description of the
${\goth g}_0^f$-module
${\goth g}_{\frac{1}{2}}$  are as
follows:

$I$. ${\goth g}$ is a simple Lie
algebra, and $\theta$ is the highest root.

$$
\begin{array}{| c | c | c || c | c | c | }
\hline
{\goth g} & {\goth
g}_0^f & {\goth
g}_{\frac{1}{2}} &  {\goth g} & {\goth
g}_0^f & {\goth
g}_{\frac{1}{2}} \\
\hline
s\ell_n \ (n \geq 3) &  g\ell_{n-2} &
\CZ^{n-2} \oplus \CZ^{n-2 *} & F_4 &
{sp}_6 & \Lambda_0^3 \CZ^6 \\
\hline
{so}_n \ (n \geq 5) & s\ell_2
\oplus {so}_{n-4} & \CZ^2 \otimes
\CZ^{n-4} & E_6 & s\ell_6 & \Lambda^3
\CZ^6 \\
\hline
{sp}_n \ (n \geq 2) & {sp}_{n-2}
&
\CZ^{n-2} & E_7 & {so}_{12}
&  spin_{12}
\\
\hline
G_2 & s\ell_2 & S^4 \CZ^2 & E_8 & E_7 &
56{-\rm dim }
\\
\hline
\end{array}
$$

$II$. ${\goth g}$ is a simple Lie
superalgebra but not a Lie algebra, $
{s\ell}_2$ is a simple component of ${\goth
g}_{{0}}$ and $\theta$ is the highest
root of this component. Below are all cases when
${\goth g}_0^f $ is a Lie
algebra ($m \geq 1$ and ${\goth g}_{\tfrac{1}{2}}$ is odd):
$$
\begin{array}{| c | c | c || c | c | c | }
\hline
{\goth g} & {\goth g}_0^f & {\goth g}_{\frac{1}{2}}
&  {\goth g} & {\goth g}_0^f & {\goth g}_{\frac{1}{2}} \\
\hline s\ell (2 | m) \ (m \neq 2) & g\ell_m &
\CZ^m \oplus \CZ^{m *} & D(2, 1 ;a) &
s\ell_2 \oplus
s\ell_2  &  \CZ^2 \otimes \CZ^2 \\
\hline
s\ell (2| 2)/\CZ I &  s\ell_2  & \CZ^2
\oplus
\CZ^2 & F(4) &  {so}_7 & {spin}_7 \\
\hline
{spo}(2 | m )  &
{so}_m &
\CZ^m & G(3) & G_2
& 7{-\rm dim }
\\
\hline
{osp}(4 | m )  & {sl}_2 \oplus
{sp}_m &
\CZ^2 \otimes \CZ^m & & & \\
\hline
\end{array}
$$

$III$. ${\goth g}$ is a simple Lie
superalgebra but not a a Lie algebra. The
remaining possibilities are:
$$
\begin{array}{| c | c | c | }
\hline
{\goth g} & {\goth
g}_0^f & {\goth
g}_{\frac{1}{2}}  \\
\hline
s\ell (m | n) \ (m \neq n\, ,\, m>2) &
g\ell(m-2 | n) &
\CZ^{m-2 | n} \oplus \CZ^{m-2 | n  *}  \\
\hline
s\ell (m| m)/\CC I\, (m>2) & s\ell(m-2 | m)  &
\CZ^{m-2 | m} \oplus \CZ^{m-2 | m  *} \\
\hline
{spo}(n | m ) \ (n \geq 4) &
{spo}(n-2 | m) & \CZ^{n-2 | m}
\\
\hline
{osp}(m | n ) \ (m \geq 5) &
{osp}(m-4 | n) \oplus s\ell_2 & \CZ^{m-4 |
n} \otimes \CZ^2 \\
\hline
 F(4) & D ( 2 , 1 ; 2) &
\stackrel{1}{\circ} - \otimes - \circ
\ \ ( (6 | 4){-\rm dim})
\\
\hline
 G(3) & {osp} ( 3 | 2 ) &
\stackrel{-3}{\otimes} \Longrightarrow
\stackrel{1}{0}
\ \ \ ( (4 | 4){-\rm dim}) \\
\hline
\end{array}
$$
\end{proposition}
{\it Proof.} The proof of this proposition is
straightforward by looking at all highest roots
$\theta$ of simple components of ${\goth g}_{{0}}$ and choosing an ordering for
which this $\theta$ is the highest root of ${\goth g}$.
$\Box$ \par \vspace{.2in} \noindent

All examples from Table~I (resp. Table~II) of Proposition~4.1
occur in the Fradkin--Linetsky list of quasisuperconformal
(resp. superconformal) algebras \cite{FL}, but the
 last two examples from Table~III are missing there.

One can check that in all cases of Table~II when $\goth g^f_0$ is
simple one has:
\begin{displaymath}
  \frac{1}{2} (h^\vee -h^\vee_0) =-1 \,,
\end{displaymath}
if we consider the normalization of the form $(.|.)$ which
restricts to the standard one on $\goth g^f_0$ (i.e.,~$(\alpha |
\alpha)$=2 for a long root of $\goth g^f_0$).  Then $(\theta |
\theta)=4$ for $osp (m|n)$, $= -3$ for $F(4)$, and
$=-\frac{8}{3}$ for $G(3)$.

It follows from (4.6) that the affine central charge in \cite{FL}
equals $k-1$ in all these cases, and this leads to a perfect
agreement of (4.4) with the Virasoro central charges in
\cite{FL}.

In the case of Table~I we take the usual normalization $(\theta |
\theta)=2$.  Then  (4.4) becomes
\begin{displaymath}
  c= \frac{k}{k+h^\vee} \dim \goth g - 6k + \frac{1}{4}
     (\dim \goth g -\dim \goth g^f)-\frac{11}{4} \, .
\end{displaymath}
If, in addition, $\goth g^f_0$ is simple, then $h^\vee -h^\vee_0
= 1,6,8,12,5$ and $\frac{10}{3}$ for $\goth g$ of type
$C_n,E_6,E_7,E_8,F_4$ and $G_2$, respectively, and again we  are
in agreement with the Virasoro central charge of \cite{FL}.

{\bf Remark 4.1.} Many examples of vertex algebras from
Proposition~4.1 are well known:


\begin{list}{}{}
\item $W_k (s\ell_2,e_{-\theta})$ is the Virasoro vertex
  algebra,

\item $W_k (s\ell_3 , e_{-\theta})$ is the
  Bershadsky--Polyakov algebra \cite{B},

\item  $W_k (spo (2|1), e_{-\theta})$ is the Neveu--Schwarz
  algebra,

\item  $W_k (spo (2|m), e_{-\theta})$ for $m \geq 3$ are the
  Bershadsky--Knizhnik algebras \cite{BeK},

\item $W_k (s\ell (2|1) =spo (2|2), e_{-\theta})$ is the $N=2$
  superconformal algebra,

\item $W_k (s\ell (2|2)/\CC I, e_{-\theta})$ is the $N=4$
  superconformal algebra,

\item  $W_k (spo (2|3), e_{-\theta})$ tensored with one fermion
  is the $N=3$ superconformal algebra (cf. \cite{GS}),

\item $W_k (D(2,1;a), e_{-\theta})$ tensored with four
  fermions and one boson is the big $N=4$ superconformal algebra (cf.~\cite{GS}).

\end{list}

\section{The example of $\hat{s \ell_2}$ and Virasoro algebra}
  (See \cite{KW1, FKW} for details).
\setcounter{equation}{0}
\label{sec:5}

Let $\goth g =s\ell_2$ with the invariant bilinear form $(a|b)=
{\rm tr}\,\, ab$; then $\Delta_+ =\{ \alpha \}$.  All possibilities for
$\hat{\Pi}'$ are as follows:
\begin{displaymath}
  \hat{\Pi}_{u,j} =\{ (u-j) K-\alpha \, , \, jK +\alpha \} \,
\hbox{ where } 0\leq j \leq u-1 \, , \, u \geq 1 \, .
\end{displaymath}
All possible levels of the admissible weights for
$\hat{\Pi}_{u,j}$ are rational numbers $k$ with a positive
denominator $u$ (relatively prime to the numerator) such that
$
  u(k+2)\geq 2 $.
The set of all admissible weights of such a level $k$ is
\begin{displaymath}
  \{ \Lambda_{k,j,n} = kD + \tfrac{1}{2} (n-j(k+2))\alpha |\,
     0 \leq j \leq u-1 \, , 0 \leq n \leq u(k+2)-2 \} \, .
\end{displaymath}
Such a weight is degenerate iff it is integrable with respect to
the root $K-\alpha$, which happens
iff $j=u-1$.  In particular, all such weights corresponding to $u=1$ are degenerate.

We have:  $W_k (s\ell_2 ,e_{-\alpha})$ is generated by the
Virasoro field $L(z)$.  Furthermore, by Theorem~3.2,
$H(L(\Lambda_{k,j,n}))$ is zero iff $j=u-1$.  Otherwise, $H
(L(\Lambda_{k,j,n}))=H_0 (L(\Lambda_{k,j,n}))$ is an irreducible highest weight module
over the Virasoro algebra defined by $L(z)$ (given by (2.5)),
corresponding to the parameters $p=u(k+2)$, $p'=u$ of the
so-called \emph{minimal series}:
\begin{displaymath}
 c^{(p,p')} = 1-6 \frac{(p-p')^2}{pp'} \, ,
 h^{(p,p')}_{j+1,n+1} =
 \frac{(p(j+1)-p'(n+1))^2 -(p-p')^2}{4pp'} \, .
\end{displaymath}
Here $p,p' \in \ZZ \, , \, 2 \leq p'<p$,  $gcd (p,p')=1$, $ 1\leq j+1 \leq p'-1$, $1 \leq n+1 \leq p-1$, which are precisely all minimal series Virasoro
modules.
The character formula for $ M=L(\Lambda_{k,j,n})$ plugged in
(3.3) gives the well-known characters of the minimal series
modules over the Virasoro algebra.  The vector
$\tilde{v}_{\Lambda_{k,j,n}}$ (see Remark~2.3) is the eigenvector with the lowest
$L_0$-eigenvalue (equal to $h^{(p,p')}_{j+1, n+1}$).

\section{ The Example of ${\rm spo}(2|1)^{\hat{}}$ and
  Neveu--Schwarz algebra}
\label{sec:6}
\setcounter{equation}{0}

In this section, $\goth g = {\rm spo}(2|1)$ with the invariant
bilinear for $(a|b)=\frac{1}{2} {\rm str}\,\, ab$.  This is a
$5$-dimensional Lie superalgebra with the basis consisting of odd
elements $e_{\alpha}, e_{-\alpha}$ and even elements $e_{2\alpha}
= [e_{\alpha},e_{\alpha}],
e_{-2\alpha}=-[e_{-\alpha},e_{-\alpha}]$ and $h=2[e_{\alpha},
e_{-\alpha}]$ such that $[h,e_{\alpha}]=e_{\alpha}$,
$[h,e_{-\alpha}]=-e_{-\alpha}$.  Then
$[h,e_{2\alpha}]=2e_{2\alpha}$,
$[h,e_{-2\alpha}]=-2e_{-2\alpha}$, $[e_{2\alpha},
e_{-2\alpha}]=h$, $[e_{\alpha}, e_{-2\alpha}]=e_{-\alpha}$,
$[e_{-\alpha} ,e_{2\alpha}]=e_{\alpha}$; $(e_{\alpha}
|e_{-\alpha})=(e_{2\alpha}| e_{-2\alpha})=\frac{1}{2}$, $(h|h)=1$.
We have: $h^{\vee}=3$ and $\Delta_+ = \{ \alpha ,2\alpha \}$.
The element $f=2e_{-2\alpha}$ is the only, up to conjugacy,
nilpotent even element, and then $x=\frac{1}{2}h$.

We have the charged free superfermions $\varphi_{j\alpha} =
\varphi_{j\alpha}(z)$ and $\varphi^*_{j\alpha} =
\varphi^*_{j\alpha}(z)$, $j=1,2$, and the neutral free fermion
$\Phi = \Phi (z)$ such that $[\Phi_{\lambda} \Phi]=1$ (since
$(f|[e_{\alpha},e_{\alpha}])=1$).  Hence we have:
\begin{displaymath}
  d=d(z)=-e_{\alpha}\varphi^*_{\alpha}+
  e_{2\alpha}\varphi^*_{2\alpha} - \tfrac{1}{2}:
  \varphi_{2\alpha} (\varphi^*_{\alpha})^2 :
  + \varphi^*_{2\alpha}+ \varphi^*_{\alpha} \Phi \, ,
\end{displaymath}
and the $\lambda$-brackets of $d$ with all generators of the
complex $\C (\goth g , f,k)$ are:
\begin{eqnarray*}
&&  [d_{\lambda} e_{2\alpha}] = 0 \, , \,
      [d_{\lambda}e_{\alpha}]
      =-e_{2\alpha}\varphi^*_{\alpha} \, ,\,
      [d_{\lambda}h] = e_{\alpha} \varphi^*_{\alpha}
      -2 e_{2\alpha}\varphi^*_{2\alpha} \, , \\[1ex]
 && [d_{\lambda}e_{-\alpha}] = -\tfrac{1}{2} h
      \varphi^*_{\alpha} + e_{\alpha} \varphi^*_{2\alpha} -\tfrac{k}{2}
      (\partial +\lambda) \varphi^*_{\alpha} \, , \,
  [d_{\lambda} e_{-2\alpha}] = -e_{-\alpha}\varphi^*_{\alpha}
     + h \varphi^*_{2\alpha} + \tfrac{k}{2} (\partial + \lambda)
     \varphi^*_{2\alpha} \, , \\[1ex]
 && [d_{\lambda}\varphi_{2\alpha}] = e_{2\alpha} +1 ,
      [d_{\lambda}\varphi_{\alpha}] =e_{\alpha}+e_{2\alpha}
      \varphi^*_{\alpha}-\Phi \, ,\,
  [d_{\lambda} \varphi^*_{2\alpha}] = -\tfrac{1}{2}
      (\varphi^*_{\alpha})^2,\, [d_{\lambda} \varphi^*_{\alpha}]=0,
      \,[d_{\lambda}\Phi] =\varphi^*_{\alpha} \, .
\end{eqnarray*}

Since $: \Phi \Phi :=0$, we have:
\begin{eqnarray*}
  J^{(h)}(z) = h(z) - :
      \varphi_{\alpha} \varphi^*_{\alpha}: +2:\varphi_{2\alpha}
      \varphi^*_{2\alpha}: \, , \,
  J^{(e_{-\alpha})}(z) = e_{-\alpha}(z)
       -\varphi_{\alpha} \varphi^*_{2\alpha} ,
       J^{(e_{-2\alpha})}(z) =e_{-2\alpha}(z) \, .
\end{eqnarray*}
It is not difficult to check that the following fields are closed
under $d_0$:
\begin{eqnarray*}
  G &=& \frac{2}{(k+3)^{1/2}} (J^{(e_{-\alpha})}
         + \frac{1}{2}\Phi J^{(h)}
         + \frac{k+2}{2} \partial \Phi) \, , \\[1ex]
  L &=& \frac{2}{k+3} (-J^{(e_{-2\alpha})} -\Phi
       J^{(e_{-\alpha})}+
       \frac{1}{4}:J^{(h)}J^{(h)}:+
       \frac{k+2}{4}\partial J^{(h)})
       -\frac{1}{2} : \Phi \partial \Phi : \, ,
\end{eqnarray*}
and that the field $L$ is equal to the Virasoro field, defined by
(2.5), modulo the image of $d_0$ so that they define the same field of $W_k (\goth g ,f)$.

Furthermore, a direct calculation with $\lambda$-brackets in $W_k (\goth g ,f)$ shows
that $L$ and $G$ form the Neveu--Schwarz algebra with central
charge $c$:
\begin{eqnarray}
  \label{eq:6.1}
  [L_{\lambda}L] &=& (\partial + 2\lambda)L
  +\frac{\lambda^3}{12}c \, , \, [L_{\lambda}G] =
  (\partial + \frac{3}{2}\lambda)G \, , \,
   [G_{\lambda}G] = 2L + \frac{\lambda^2}{3}c \, , \, \\
\label{eq:6.2}
c &=& \frac{3}{2} \left( 1-\frac{2(k+2)^2}{k+3} \right) \, .
\end{eqnarray}

The set of positive roots of $\hat{\goth g} $ is $(n \in \ZZ)$:
\begin{displaymath}
  \hat{\Delta}_+ = \{ nK |n>0 \} \cup
  \{ j \alpha + nK | n \geq 0 \, , \, j=1,2 \} \, , \
  \cup \{ -j\alpha +nK | n>0 \, , \, j=1,2 \} \, ,
\end{displaymath}
and the set of simple roots is
\begin{displaymath}
  \hat{\Pi} =\{ \alpha_0 =K-\alpha \, , \,
  \alpha_1 =2\alpha \} \, .
\end{displaymath}
All possibilities for the sets $\hat{\Pi'}$ of simple roots of
subsets $\hat{\Delta}'_+$ of $\hat{\Delta}_+$ that are
isomorphic to a set of positive roots of an affine superalgebra,
are of three types:  the \emph{principal} ones (isomorphic to
$\hat{\Pi}$), the \emph{even type} ones, isomorphic to the set
of simple roots of type~$A^{(1)}_1$, and the \emph{subprincipal}
ones, isomorphic to the set of simple roots of the twisted
affine superalgebra $C^{(2)} (2) $ \cite{K2}.

All admissible weights for $\hat{\goth g}$ are of the form:
\begin{displaymath}
  \Lambda_{k,j,n} =2(n-j (k+3)) \Lambda_0
  + (\tfrac{1}{2} k-n + j(k+3)) \Lambda_1 \, ,
\end{displaymath}
where $\Lambda_0,\Lambda_1$ are the fundamental weights,
$k=\frac{v}{u} \in \QZ$ is the level $(u,v \in \ZZ\, , \, u \geq
1\, , \, {\rm gcd}  (u,v)=1)$, and $j,n \in \frac{1}{2} \ZZ_+$.  The
ranges of $k$ and $j,n$ are described below.

All principal admissible weights have level $k$ such that its
denominator $u$ is a (positive) odd integer, $v$ is an even
integer, and $u(k+3) \geq 3$.  Both $j,n$ are integers satisfying
the following conditions:
\begin{eqnarray*}
  \begin{array}{lclc}
\hbox{(i)} & 0 \leq j \leq \frac{u-1}{2} & \hbox{and} &
0 \leq n \leq \frac{u(k+3)-3}{2} \, \hbox{ \quad  or}\\[1ex]
\hbox{(ii)} & \frac{u+1}{2}\leq j \leq u-1 & \hbox{and} &
\frac{u(k+3)+1}{2}\leq n \leq u(k+3)-1 \, .
  \end{array}
\end{eqnarray*}

In case~(i), $\hat{\Pi}' =\{ jK + \alpha_0 \, , \,
(u-1-2j)K+\alpha_1 \}$.  Hence, by Theorem~3.2, the principal
admissible weight $\Lambda_{k,j,n}$ is degenerate iff $j=\frac{u-1}{2}$.

In case~(ii), $\hat{\Pi}'=\{ (u-j)K-\alpha_0 \, , \, (2j+1-u)
K-\alpha_1 \}$ and all the admissible weights are non-degenerate.

For the even type admissible weights, $u$ is even and $v$ is odd,
and $u(k+3)\geq 2$.  Both $j \, , \, n \in \frac{1}{2}+\ZZ$ and satisfy
the inequalities:  $0 <j\leq u-\frac{1}{2}$, $0<n<u(k+3)-1$.  In
this case $\hat{\Pi}=\{ (2j+1)K-\alpha_1 \, , \,
(2(u-j)-1)K+\alpha_1 \}$, and $\Lambda_{k,j,n}$ is degenerate iff
$j=u-\frac{1}{2}$.

For the subprincipal admissible weights, both $u$ and $v$ are odd
integers, and $u(k+3)\geq 1$.  Both $j,n$ are integers,
satisfying the inequalities:  $0 \leq j \leq u-1 \, , \, 0\leq n
\leq u(k+3)-1$.  In this case $\hat{\Pi}' =\{ jK +\alpha_0 \, ,
\, (u-j)K-\alpha_0 \}$ and all the admissible weights are
non-degenerate.

The characters of all admissible $spo (2|1)^{\hat{}}\,$-modules are
known \cite{KW1}.  Applying to them Theorem~3.1 we obtain
the well known characters of all minimal series modules of the
Neveu-Schwarz algebra (see e.g.~\cite{KW1}).

Recall that these minimal series correspond to central charges
equal
\begin{equation}
  \label{eq:eq:6.3}
  c^{(p,p')} =\frac{3}{2} \left(1-\frac{2(p-p')^2}{pp'}\right) \, ,
\end{equation}
where $p,p' \in \ZZ$, $2 \leq p' <p$, $p-p' \in 2\ZZ$, $gcd
\left( \frac{p-p'}{2} \, , \, p' \right) =1$, and the minimal
eigenvalue of $L_0$ equals
\begin{equation}
  \label{eq:6.4}
  h^{(p,p')}_{r,s} =\frac{(pr-p's)^2-(p-p')^2}{8pp'} \, ,
\end{equation}
where $r,s \in \ZZ$, $1 \leq r \leq p'-1$, $1\leq s\leq p-1$,
$r-s \in 2\ZZ$.  The corresponding normalized character is as follows:
\begin{equation}
  \label{eq:6.5}
  \chi^{(p,p')}_{r,s} (\tau) = \frac{1}{\eta_{1/2}(\tau)}
  (\theta_{\frac{pr-p's}{2} \, , \, \frac{pp'}{2}} (\tau)
  -\theta_{\frac{pr+p's}{2} \, , \, \frac{pp'}{2}} (\tau)) \, ,
\end{equation}
where $\eta_{1/2}(\tau) = \frac{\eta (\tau /2) \eta (2\tau)}{\eta
  (\tau)}$ and $\theta_{n,m} (\tau) =\sum_{k \in \ZZ
  +\frac{n}{2m}} e^{2\pi imk^2 \tau}$.  Another way of writing
these characters, via the Weyl group $\hat{W}$ of $\hat{\goth g}$,
is as follows:
\begin{equation}
  \label{eq:6.6}
  \chi^{(p,p')}_{r,s} (\tau) = \frac{1}{\eta_{1/2}(\tau)}
     \sum_{w \in \hat{W}} \epsilon (w) {q^{\frac{pp'}{4} |
       \frac{w(\Lambda + \hat{\rho})}{p}-
       \frac{\Lambda'+\hat{\rho}}{p'}|^2}} \, ,
\end{equation}
where $\Lambda +\hat{\rho} = p\Lambda_0 +s \frac{\alpha_1}{2}$,
$\Lambda'+\hat{\rho} =p'\Lambda_0 +r\frac{\alpha_1}{2}$, $1 \leq s\leq
p-1$, $1 \leq r \leq p'-1$.

In the principal case we let $p=u(k+3)$, $p'=u$.  Then
(\ref{eq:6.2}) becomes $c=c^{(p,p')}$, given by (6.3).  Using
Theorem~3.1 and (\ref{eq:6.6}) we obtain:
\begin{eqnarray*}
  q^{-c^{(p,p')}/24} {\rm ch}_{H(L(\Lambda_{k,j,n}))} =
\left\{
    \begin{array}{llll}
      \chi^{(p,p')}_{p'-2j-2, p-2n-2} (\tau)
            \hbox{   in case (i)}\\[2ex]
       \chi^{(p,p')}_{2j-p',2n-p}(\tau)
             \hbox{  in case (ii)}
    \end{array} \right. \, ,
\end{eqnarray*}
so we get all characters of minimal series for which both $p$ and
$p'$  are odd.

In the even type cases and subprincipal cases we let $p=2u(k+3)$,
$p'=2u$.  Then again (\ref{eq:6.2}) becomes $c=c^{(p,p')}$, and we
obtain
\begin{displaymath}
  q^{-c^{(p,p')}/24} {\rm ch}_{H(L(\Lambda_{k,j,n}))} =
  \chi^{(p,p')}_{2p'-2j-1,2p-2n-1} \, ,
\end{displaymath}
so we get all characters of minimal series for which both $p$ and
$p'$ are even.  Both $r$ and $s$ are either even  (in the even
type case) or odd (in the subprincipal case).

\section{The example of $s\ell (2|1)^{\hat{}}$ and $N=2$
  superconformal algebra}
\label{sec:7}
\setcounter{equation}{0}

In this section, $\goth g =s\ell (2|1)$ with the invariant
bilinear form $(a|b) = {\rm str} \,\, ab$.  This is the Lie
superalgebra of traceless matrices in the superspace $\CC^{2|1}$
whose even part is $\CC \epsilon_1 +\CC \epsilon_3$ and odd part
is $\CC \epsilon_2$, where $\epsilon_1, \epsilon_2,\epsilon_3$ is
the standard basis.  We shall denote by $E_{ij}$ the standard
basis of the space of matrices.  We shall work in the following
basis of $\goth g$:
\begin{eqnarray*}
&&  e_1 = E_{12} \, , \, e_2=E_{23} \, , \, e_{12}=-E_{13} \, , \,
  f_1=E_{21} \, , \, f_2=-E_{32} \, , \\
&&  f_{12} = -E_{31} \, , \, h_1=E_{11}+E_{22} \, , \,
  h_2 =-E_{22}-E_{33} \, .
\end{eqnarray*}
The elements $e_i,f_i,h_i \,\, (i=1,2)$ are the Chevalley
generators of $\goth g$ \cite{K1}.  Elements $e_i,f_i \,\,
(i=1,2)$ are all odd elements of $\goth g$.  We pick the Cartan
subalgebra $\goth h = \CC h_1 + \CC h_2$.  The simple roots
$\alpha_1$ and $\alpha_2$ are the roots attached to $e_1$ and
$e_2$, and $\Delta_+ =\{ \alpha_1 \, , \, \alpha_2 \, , \,
\alpha_1 + \alpha_2 \}$.  We have: $\alpha_i = h_i$ $(i=1,2)$ (under the identification of $\goth h$ with $\goth h^*$).

Since $\goth g_{\bar{0}} =\CC e_{12} +\CC f_{12} +\goth h \,\,
(\simeq g\ell_2)$, there is only one, up to conjugacy, nilpotent
element $f=f_{12}$, which embeds in the following
$s\ell_2$-triple $\langle e=e_{12}, x=\frac{1}{2}(h_1+h_2),f
\rangle$.  The corresponding $\frac{1}{2}\ZZ$-gradation looks as follows:
\begin{displaymath}
  \goth g = \CC f \oplus (\CC f_1 + \CC f_2) \oplus \goth h
     \oplus (\CC e_1 + \CC e_2) \oplus \CC e \, .
\end{displaymath}
We have:  $\goth g^f = \CC f + \CC f_1 + \CC f_2 +\CC
(h_1-h_2)$.  There is only one other good
$\frac{1}{2}\ZZ$-gradation (which is non-Dynkin).  It will be
considered after the discussion related to the Dynkin gradation
is completed.

We have three pairs of charged free fermions:  $\varphi_1 =
\varphi_1 (z) \, , \, \varphi^*_1 = \varphi^*_1 (z) \, , \,
\varphi_2 = \varphi_2 (z) \, , \, \varphi^*_2=\varphi^*_2 (z)$
(which are even fields), and $\varphi_{12}=\varphi_{12}(z) \, ,
\, \varphi^*_{12} = \varphi^*_{12}(z)$ (which are odd fields).
There are two neutral free fermions:  $\Phi_i = \Phi_i (z) \,\,
(i=1,2)$,  they are odd, and their $\lambda$-bracket is easily
seen to be:
\begin{displaymath}
  [\Phi_{i\lambda} \Phi_j] =-1 \hbox{  if  } i\neq j, =0 \hbox{  otherwise.}
\end{displaymath}
Hence the field $d=d(z)$ is as follows:
\begin{displaymath}
  d=-e_1\varphi^*_1 - e_2 \varphi^*_2 + e_{12} \varphi^*_{12}
  + \varphi_{12} \varphi^*_1 \varphi^*_2 + \varphi^*_{12}
  + \varphi^*_1 \Phi_1 + \varphi^*_2 \Phi_2 \, .
\end{displaymath}
Its $\lambda$-brackets with the generators of the complex $C
(\goth g ,f,k) $ are as follows:
\begin{eqnarray*}
 &&  [d_{\lambda}e_1] = e_{12}\varphi^*_2 \, , \,
     [d_{\lambda} e_2] = e_{12} \varphi^*_1 \, , \,
     [d_{\lambda} e_{12}] =0 \, , \\
 && \left[d_{\lambda}f_1\right] = -h_1 \varphi^*_1 - e_1 \varphi^*_{12}
     - (\partial + \lambda) k\varphi^*_1 \, , \\
 && \left[d_{\lambda}f_2\right] = h_2 \varphi^*_2 -e_1 \varphi^*_{12}
     -(\partial + \lambda) k\varphi^*_2 \, , \\
&&  \left[d_{\lambda}f_{12}\right] = f_2 \varphi^*_1 + f_1 \varphi^*_2
     + (h_1 + h_2) \varphi^*_{12} + (\partial +\lambda)
     k\varphi^*_{12} \, , \\
 && \left[d_{\lambda}h_1\right] = e_2 \varphi^*_2 -e_{12}\varphi^*_{12} \, ,
     \, [d_{\lambda} h_2]=e_1\varphi^*_1-e_{12}\varphi^*_{12} \, ,\\
 && \left[d_{\lambda}\varphi_1\right] = e_1 -\varphi_{12}\varphi^*_2
     -\Phi_1 \, , \, [d_{\lambda} \varphi_2] =e_2-\varphi_{12}
     \varphi^*_1 -\Phi_2 \, , \\
 && \left[d_{\lambda}\varphi_{12}\right] = e_{12} +1 \, , \,
     [d_{\lambda} \varphi^*_j] =0 \, , \,
     [d_{\lambda} \varphi^*_{12}] =\varphi^*_1 \varphi^*_2 \, , \\
 && \left[d_{\lambda}\Phi_1\right] = -\varphi^*_2 \, , \,
  [d_{\lambda}\Phi_2] = -\varphi^*_1 \, .
\end{eqnarray*}

Furthermore, we have the fields:
\begin{eqnarray*}
 && J^{(h_1)} (z) = h_1 (z) -:\varphi_2\varphi^*_2 :+ :
     \varphi_{12} \varphi^*_{12}: \, ,\\
 && J^{(h_2)} (z) = h_2 (z) -: \varphi_1 \varphi^*_1 :
  + : \varphi_{12} \varphi^*_{12}: \, ,\\
 && J^{(f_1)} (z)= f_1 (z)  +: \varphi_2 \varphi^*_{12}:\, ,\,\,
     J^{(f_2)}(z) =f_2 (z) +:\varphi_1 \varphi^*_{12}: \, , \,
     J^{(f_{12})} (z) =f_{12}(z) \, .
\end{eqnarray*}

One easily calculates the $\lambda$-brackets of $d$ with these
fields, using (2.4):
\begin{eqnarray*}
 && \left[ d_{\lambda} J^{(h_1)} \right] =
     \varphi^*_{12} + \varphi^*_2 \Phi_2 \, , \,
     [d_{\lambda} J^{(h_2)}] = \varphi^*_{12}
     + \varphi^*_1 \Phi_1 \, , \\
  && \left[ d_{\lambda} J^{(f_1)} \right] =
     -:\varphi^*_1 J^{(h_1)}:+\varphi^*_{12}\Phi_2
     -(k+1)(\partial +\lambda) \varphi^*_1 \, , \\
 &&  \left[ d_{\lambda} J^{(f_2)} \right] =
     -:\varphi^*_2 J^{(h_2)}:+: \varphi^*_{12} \Phi_1 :
     -(k+1)(\partial + \lambda) \varphi^*_2 \, , \\
 &&  \left[ d_{\lambda} J^{(f_{12})} \right]=
     :\varphi^*_1 J^{(f_2)}:+: \varphi^*_{2}
     J^{(f_1)} :+: \varphi^*_{12} J^{(h_1+h_2)}:
     +k (\partial + \lambda) \varphi^*_{12} \, .
\end{eqnarray*}

Using this, one checks directly that the following fields are
closed under $d_0$:
\begin{eqnarray*}
  J &=& J^{(h_1-h_2)} +: \Phi_1 \Phi_2 : \, ,\\
  L &=& - \frac{1}{k+1} (J^{(f_{12})} +:\Phi_1 J^{(f_1)}
        :+: \Phi_2 J^{(f_2)} :-: J^{(h_1)} J^{(h_2)} :) \\
    && + \tfrac{1}{2} (\partial J^{(h_1+h_2)}
       +:\Phi_1 \partial \Phi_2:+:\Phi_2 \partial \Phi_1:) \, , \\
  G^+ &=& - \frac{1}{k+1} (J^{(f_1)} -: \Phi_2 J^{(h_1)}:)
        + \partial \Phi_2 \, , \\
  G^- &=& J^{(f_2)} -: \Phi_1 J^{(h_2)}:-(k+1)\partial \Phi_1 \, .
\end{eqnarray*}
Moreover, one can show that the field $L$ coincides with the
Virasoro field defined by (2.5), modulo to the image of $d_0$, and therefore they give the same field of $W_k (\goth g ,f)$.

A direct calculation with $\lambda$-brackets shows that $J$, $L$,
$G^+$ and $G^-$ form the $N=2$ superconformal algebra with
central charge $c=-3(2k+1)$:
\begin{eqnarray}
\nonumber
&&  [L_{\lambda}L] = (\partial +2\lambda)L+\lambda^2
     \frac{c}{12} \, , \, [J_{\lambda}J] =\lambda
     \frac{c}{3} \, , \\[1ex]
\label{eq:7.1}
&&  [{G^{\pm}}_{\lambda} G^{\pm}] = 0 \, , \,
     [J_{\lambda} G^{\pm}] = \pm G^{\pm} \, , \,
  [{G^+}_{\lambda}G^-] = L+\frac{1}{2} (\partial +2\lambda)
     J+\lambda^2 \frac{c}{6} \, , \\[1ex]
\nonumber
&&  [L_{\lambda}J] = (\partial + \lambda)J \, , \,
     [L_{\lambda}G^{\pm}] =
     (\partial + \tfrac{3}{2} \lambda)G^{\pm} \, .
\end{eqnarray}

The good non-Dynkin $\frac{1}{2}\ZZ$-gradation looks as follows:
\begin{displaymath}
  \goth g = (\CC f_{12}+\CC f_2) \oplus 0 \oplus
  (\CC e_1 + \CC f_1 + \goth h) \oplus 0 \oplus
  (\CC e_{12}+\CC e_2) \, .
\end{displaymath}
It corresponds to $x=h_1$.  As before, we take $f=f_{12}$.

In this case we have two pairs of charged free fermions.  Hence
the field $d=d(z)$ is as follows:
\begin{displaymath}
  d=-e_2 \varphi^*_2 + e_{12}\varphi^*_{12} + \varphi^*_{12}\, ,
\end{displaymath}
and its $\lambda$-brackets with the generators of the complex are
as follows:
\begin{eqnarray*}
 &&  [d_{\lambda}e_1] = e_{12}\varphi^*_2 \, , \,
     [d_{\lambda} e_2]=0 \, , \, [d_{\lambda}e_{12}]=0 \, , \\
  && [d_{\lambda}h_1] = e_2 \varphi^*_2-e_{12}\varphi^*_{12} \, ,\,
     [d_{\lambda}h_2]=-e_{12}\varphi^*_{12} \, , \\
   &&[d_{\lambda }f_1] = -e_2\varphi^*_{12} \, , \,
     [d_{\lambda}f_2]=-h_2 \varphi^*_2 + (h_1+h_2)
     \varphi^*_{12} -k(\partial + \lambda)\varphi^*_2 \, , \\
   &&[d_{\lambda}f_{12}] = f_1\varphi^*_2 + (h_1 + h_2)
     \varphi^*_{12} + k (\partial+ \lambda) \varphi^*_{12} \, ,\\
   &&[d_{\lambda}\varphi_2] = e_2 \, , \,
     [d_{\lambda}\varphi^*_2] =0 \, , \,
     [d_{\lambda}\varphi_{12}] =e_{12} +1 \, , \,
     [d_{\lambda}\varphi^*_{12}]=0 \, .
\end{eqnarray*}

Furthermore, we have the fields:
\begin{eqnarray*}
&& J^{(e_1)}(z) = e_1 (z) -\varphi_{12}\varphi^*_2 \, ,\,
  J^{(h_1)}(z) = h_1 (z) -: \varphi_2 \varphi^*_2 :+:
      \varphi_{12} \varphi^*_{12}: \, ,\, \\
  && J^{(h_2)}(z) = h_2(z) +:\varphi_{12}\varphi^*_{12}:\, ,\\
  && J^{(f_1)}(z) = f_1(z) +:\varphi_2\varphi^*_{12}: \, ,
     J^{(f_2)}(z)=f_2(z) \, , \, J^{(f_{12})} (z)
     =f_{12}(z) \, .
\end{eqnarray*}

One easily calculates the $\lambda$-brackets of $d$ with these
fields, using (2.4):
\begin{eqnarray*}
&&  [d_{\lambda}J^{(h_1)}]= [d_{\lambda},J^{(h_2)}]
     =\varphi^*_{12} \, , \\
{}&&  [d_{\lambda}J^{(e_1)}] = -\varphi^*_2 \, , \,
     [d_{\lambda} J^{(f_1)}] =0 \, , \\
{}&&  [d_{\lambda}J^{(f_2)}] = -:\varphi^*_2 J^{(h_2)}:
     +:\varphi^*_{12} J^{(e_1)}:-k(\partial +\lambda)
     \varphi^*_2 \, , \\
{} &&    [d_{\lambda}J^{(f_{12})}]=  :\varphi^*_{12}J^{(h_1 +h_2)}:
     + :\varphi^*_2 J^{(f_1)}:+k(\partial +\lambda)
     \varphi^*_{12} \, .
\end{eqnarray*}

Using this one checks directly that the following fields are
closed under $d_0$:
\begin{eqnarray*}
&&  J = J^{(h_1-h_2)} \, , \\
&&  L'= -\frac{1}{k+1} \left(J^{(f_{12})} + :J^{(e_1)}
       J^{(f_1)}: -:J^{(h_1)}J^{(h_2)}: \right)
       + \tfrac{1}{2} \partial J^{(h_1 +h_2)} \, , \\
&&  G^+ = -\frac{1}{k+1} J^{(f_1)} \, , \,
     G^- =J^{(f_2)}-:J^{(h_2)} J^{(e_1)}: -k\partial
      J^{(e_1)} \, .
\end{eqnarray*}

A direct calculation with $\lambda$-brackets shows that $J$, $L'$,
$G^+$ and $G^-$ form the $N=2$ superconformal algebra with
central charge $c=-3(2k+1)$.  However, in this case the relation
between $L'$ and the field $L$, defined by (2.5), is more
complicated.  One can show that in $W_k (\goth g, x,f)$ one has:
\begin{equation}
\label{eq:7.2}
  L=L' + \frac{1}{2} \partial J \, .
\end{equation}
The four fields $J$, $L$, $G^+$ and $G^-$ form the Ramond type
basis of $N=2$ superconformal algebra (\cite{RY,R}):
\begin{eqnarray}
\nonumber
&&   [L_{\lambda}L]= (\partial +2\lambda) L \, , \,
       [J_{\lambda} J]=\lambda \frac{c}{3} \, ,\\[1ex]
\label{eq:7.3}
&& {} [{G^{\pm}}_{\lambda} G^{\pm}]=0 \, , \,
      [J_{\lambda}G^{\pm}] = \pm G^{\pm} \, , \,
      [{G^+}_{\lambda} G^-] = L+\lambda J-\lambda^2 \frac{c}{6}\, , \\[1ex]
\nonumber
&& {} [L_{\lambda}J]= (\partial +\lambda)J-\lambda^2
       \frac{c}{6} \, , \, [L_{\lambda}G^+]=(\partial +\lambda)
       G^+ \, , \, [L_{\lambda}G^-]=(\partial +2\lambda)G^-\, .
\end{eqnarray}

The set of positive roots of the affine superalgebra $\hat{\goth
  g} =s\ell (2|1)^{\hat{}}$ is $(n \in \ZZ)$:
\begin{eqnarray*}
  \hat{\Delta}_+ &=& \{ nK \hbox{  of multiplicity  } 2| n>0 \}\\
  && \cup \{ \alpha + nK | \alpha \in \Delta_+ \, , \,
    n \geq 0 \} \,\, \cup \{ -\alpha +nK |
    \alpha \in \Delta_+ \, , \,  n>0 \} \, ,
\end{eqnarray*}
and the set of simple roots is
\begin{displaymath}
  \hat{\Pi} = \{ \alpha_0 = K-\alpha_1 -\alpha_2 \, , \, \,
     \alpha_1 \, , \, \alpha_2 \} \, .
\end{displaymath}
All admissible subsets $\hat{\Delta}'_+$ of $\hat{\Delta}_+$ are
principal, and the corresponding sets of simple roots are as follows:
\begin{eqnarray*}
  \hat{\Pi}_b &=& \{ b_0 K+ \alpha_0 \, , \,
      b_1 K + \alpha_1 \, , \, b_2 K + \alpha_2 \} \, ,
      \hbox{ where } b=(b_0,b_1,b_2) \in \ZZ^3_+ \, ,\\
      \hat{\Pi}^-_b &=& \{ b_0 K -\alpha_0 \, , \,
      b_1 K-\alpha_1 \, , \, b_2 K-\alpha_2 \} \, ,
      \hbox{ where } b=(b_0,b_1,b_2) \in (1+\ZZ_+)^3 \, .
    \end{eqnarray*}

For the set $\hat{\Pi_b}$, the boundary admissible weights $\Lambda$ are determined from the equation
\begin{equation}
  \label{eq:7.4}
  (\Lambda +\hat{\rho} | b_0 K +\alpha_0) =1 \, , \,
  (\Lambda + \hat{\rho} | b_1 K +\alpha_1) =
  (\Lambda +\hat{\rho} |b_2 K +\alpha_2)=0 \, .
\end{equation}
Adding these equations, we get $(\Lambda +\hat{\rho} | uK)=1$, where $u=b_0+b_1+b_2+1$.  Since $(\hat{\rho}|K)=1$, we obtain that the level of $\Lambda$ is given by
\begin{equation}
  \label{eq:7.5}
  k=\frac{1}{u}-1 \, , \hbox{ where }
  u=b_0+b_1+b_2+1 \, ,
\end{equation}
and from (\ref{eq:7.4}) we obtain:  $(\Lambda |\alpha_i)=-\frac{b_i}{u}$, $i=0,1,2$.  Hence, denoting by $\Lambda_i$ $(i=0,1,2)$ the fundamental weights, i.e.,~$(\Lambda_i|\alpha_j)=\delta_{ij} \, , \, (\Lambda_i |D) =0$, we obtain the unique boundary admissible weight corresponding to $\hat{\Pi}_b$:
\begin{displaymath}
  \Lambda_b =-\frac{1}{u} (b_0 \Lambda_0 + b_1 \Lambda_1
     +b_2 \Lambda_2) \, , \, u=b_0+b_1+b_2+1 \, .
\end{displaymath}
It is easy to see that this weight is nondegenerate iff $b_0 \geq 1$, which we will assume.

Recall that $\goth h^f = \CC (h_1-h_2)$.  We let in (3.3) $h=z(h_1-h_2)$, $z \in \CC$, and let $y=e^{2\pi iz}$.  We shall calculate the normalized Euler--Poincar\'e character
\begin{equation}
  \label{eq:7.8}
  \chi_{H(M)} (\tau ,z):= q^{-c/24} {\rm ch}_{H(M)}
     (z (h_1-h_2)) \, ,
\end{equation}
where $c$ is the central charge (given by formula (\ref{eq:7.9}) below).

The conjectural character formula (3.5) gives in this case:
\begin{equation}
  \label{eq:7.7}
  \hat{R} {\rm ch}_{L(\Lambda_b)} = e^{\Lambda_b}
     \Pi^{\infty}_{j=1}
     \frac{(1-q^{u(j-1)+b_0} e^{-\alpha_0})
          (1-q^{u j-b_0} e^{\alpha_0}) (1-q^j)^2}
        {(1+q^{u(j-1)+b_1} e^{-\alpha_1})
          (1+q^{u_j-b_1} e^{\alpha_1})
          (1+q^{u(j-1)+b_2} e^{-\alpha_2})
          (1+q^{uj-b_2} e^{-\alpha_2})} \, .
\end{equation}

Due to (3.3), $\chi_{H(L(\Lambda_b))}$ is obtained from this formula in the case of the Dynkin gradation by the specialization
\begin{equation}
   \label{eq:7.8}
   e^{-\alpha_0} =1 \, , \, e^{-\alpha_1} = y q^{\frac{1}{2}} \, , \,
      e^{-\alpha_2} = y^{-1}q^{\frac{1}{2}}
\end{equation}
(and multiplication by the specialized product).  In order to write down the explicit formula, it is convenient to introduce the following important function:
\begin{displaymath}
  F(\tau ,z_1,z_2) = \Pi^{\infty}_{n=1}
  \frac{(1-q^n)^2 (1-e^{-2\pi i (z_1+z_2)} q^n)
         (1-e^{2\pi i (z_1+z_2)} q^{n-1}) }
       {(1-e^{-2\pi i z_1} q^n) (1-e^{-2\pi i z_1} q^{n-1})
         (1-e^{-2\pi i z_2} q^n)(1-e^{2\pi i z_2}q^{n-1})}
\end{displaymath}
and the following its specializations:
\begin{displaymath}
  F^{(u)}_{j,\ell} (\tau ,z) =q^{\frac{j\ell}{u}}
  e^{\frac{2\pi i (j-\ell)z}{u}}
  F(u \tau \, , \, j\tau - z-\tfrac{1}{2} \, , \,
  \ell \tau +z+\tfrac{1}{2}) \, .
\end{displaymath}

Note that plugging (\ref{eq:7.5}) in the formula for the central charge $c=-3(2k+1)$, we obtain:
\begin{equation}
  \label{eq:7.9}
  c=3-\frac{6}{u} \, , \, u=2,3, \ldots \, .
\end{equation}
This is precisely the central charge of the minimal series
representations of the $N=2$ superconformal algebra.  Recall that all
these representations with given central charge (\ref{eq:7.9}) are
parameterized by a pair of numbers $j,\ell \in \frac{1}{2} \ZZ$
satisfying inequalities $0<j,\ell ,j+\ell <u$, the minimal eigenvalue
of $L_0$ being $\frac{j\ell-1/4}{u}$ and the corresponding eigenvalue of
$J_0$ being $\frac{j-\ell}{u}$.

The specialization (\ref{eq:7.8}) of the right-hand side of
(\ref{eq:7.7}) gives $F_{b_1+\frac{1}{2}, b_2+\frac{1}{2}} \,\, (\tau
,z)$, and the specialization in (\ref{eq:7.8}) of the product in (3.3)
gives $F^{(2)}_{\frac{1}{2}, \frac{1}{2}} (\tau ,z)^{-1}$.  Hence,
letting $j=b_2+\frac{1}{2}$ and $\ell = b_1+\frac{1}{2}$,
formula~(3.3) gives the well known (normalized) characters of the
minimal series of $N=2$ superconformal algebra (cf.~\cite{D,Ki,M}):
\begin{equation}
  \label{eq:7.10}
  \chi_{H(L(\Lambda_b))} (\tau ,z) =\chi^{(u)}_{j,\ell}
     (\tau ,z) := F^{(u)}_{j,\ell} (\tau ,z) /
     F^{(2)}_{\frac{1}{2}, \frac{1}{2}} (\tau ,z) \, .
\end{equation}
Note that, given $u \geq 2$, the range of $j$ and $\ell$ exactly
corresponds to the range of $b_1$ and $b_2$ (defined by
(\ref{eq:7.5})), since $b_0 \geq 1$.  It is also easy to see that
(2.19) for $\Lambda =\Lambda_b$ gives the minimal eigenvalue of $L_0$,
and the corresponding eigenvalue of $J_0$ is indeed
$\Lambda_{b}(h_1-h_2)$.  Using Remark~2.3, one can conclude that $H_0
(L(\Lambda_b)) \neq 0$ (if $\Lambda_b$ is non-degenerate).  Hence, by
Conjecture~3.3B, $H_j (L(\Lambda_b))=0$ for $j \neq 0$, and therefore
$H_0 (L(\Lambda_b))$ is the irreducible module of minimal series
corresponding to the parameters $u,j,\ell$.

In a similar fashion, for $\Pi^-_b$ the only boundary admissible weight is
\begin{displaymath}
  \Lambda^-_b = \left( \frac{b_0}{u} -2 \right) \Lambda_0 +
  \frac{b_1}{u} \Lambda_1 +\frac{b_2}{u} \Lambda_2 \, , \,
  u=b_0+b_1+b_2-1 \, .
\end{displaymath}
All these weights are non-degenerate.

In a similar fashion, $\chi_{H(L(\Lambda^-_b))}$ is obtained from
(3.3) by using (\ref{eq:7.7}) and the specialization (\ref{eq:7.8}).
It turns out that we again recover all characters of the $N=2$ minimal
series (\ref{eq:7.10}), where we set $j=b_1-\frac{1}{2}$, $\ell =b_2
-\frac{1}{2}$.  All other statements made about $\Lambda_b$ hold for
$\Lambda^-_b$ as well.

We proceed in exactly the same way in the case of a non-Dynkin gradation.  In this case the specialization (\ref{eq:7.8}) is replaced by
\begin{displaymath}
  e^{-\alpha_0} =1 \, , \, e^{-\alpha_1}=y \, , \,
  e^{-\alpha_2} =qy^{-1} \, .
\end{displaymath}
In a similar fashion we recover all Ramond type characters of the $N=2$ superconformal algebra (meaning that we use the Virasoro field from the Ramond type basis (\ref{eq:7.2}), cf. \cite{RY}, \cite{R}):
\begin{equation}
\label{eq:7.11}
  e^{-\pi icz} {\rm ch}_{H(L(\Lambda_b))} = \chi^{(u)R}_{j,\ell}
  (\tau ,z) := F^{(u)}_{j,\ell} (\tau ,z)/F^{(2)}_{1,0}
  (\tau ,z) \, ,
\end{equation}
where $j=b_2+1$ and $\ell =b_1$ so that the range of $j,\ell$ is exactly right: %
\begin{displaymath}
0<j \, , \, j+\ell <u \, , \, 0\leq \ell <u\, .
\end{displaymath}
Likewise, the same result holds for $\Lambda^-_b$ if we let $j=b_1$, $\ell =b_2-1$.  (Incidentally, using $L'$ instead of $L$, see (\ref{eq:7.2}), we get again $\chi^{(u)}_{j,\ell}$.)

Note that for the Ramond type basis (\ref{eq:7.3}) the fields $G^+$
and $G^-$ have conformal weights $1$ and $2$, respectively.  Letting
$G^+ (z) = \sum_{n \in \ZZ} G^+_n z^{-n-1}$, $G^- (z)= \sum_{n \in \ZZ}
  G^-_n z^{-n-2}$, and introducing the constant term corrections:
$\tilde{L} (z) = L(z) + \frac{c}{24z^2}$, $\tilde{J} (z) =J(z)
-\frac{c}{6z}$, formula (\ref{eq:7.3}) gives us exactly the
commutation relation of the Ramond type $N=2$ superalgebra.  Using
$\tilde{L}_0$ and $\tilde{J}_0$ in place of $L_0$ and $J_0$ in the
definition of the normalized Euler--Poincar\'e character, the
definition (\ref{eq:7.11}) turns into the standard
definition~(7.6).

Recall \cite{RY}, \cite{KW3}, that, given $u$, the span of all $N=2$
characters, Ramond type characters and the corresponding
supercharacters (obtained, up to a constant factor, by replacing
$\tau$ by $\tau +1$ in the character) form the minimal
$SL_2(\ZZ)$-invariant subspace containing the ``vacuum'' character
$\chi^{(u)}_{\frac{1}{2}, \frac{1}{2}}$.  Thus, taking quantum
reduction for all good gradations of $s \ell (2|1)$ of all boundary
admissible highest weight $s\ell (2|1)^{\hat{}}$-modules, we get an
$SL_2 (\ZZ)$-invariant space spanned by all characters and
supercharacters.

\section*{ Acknowledgments. }
We would like to thank ESI, Vienna, where we began this work in the
summer of 2000, MSRI, Berkeley, where this work was continued in the
spring of 2002, and M.I.T., where this paper was completed in the fall
of 2002, for their hospitality.  This paper was partially supported by NSF
grants DMS9970007 and DMS0201017, NSC grant 902115M001020 of Taiwan,
and grant in aid 13440012 for scientific research Japan.

\end{document}